\pdfoutput=1

\documentclass[12pt, leqno, fleqn]{article}

\usepackage{amssymb,amsmath,amsfonts,eurosym,geometry,ulem, color,setspace,sectsty,comment,footmisc,caption,pdflscape,array,hyperref, graphicx}
\usepackage{booktabs,siunitx, makecell}
\usepackage{natbib}
\usepackage{mathtools}
\usepackage{breqn}
\usepackage{cellspace}
\usepackage{makecell}
\usepackage{chngcntr}
\usepackage{endnotes}
\usepackage{subcaption}
\usepackage{threeparttable}

\DeclareMathOperator*{\argmin}{arg\,min}

\usepackage{threeparttable}

\usepackage{dcolumn}
\newcolumntype{d}[1]{D..{#1}}

\usepackage{subcaption}

\usepackage{arydshln} 
\usepackage[T1]{fontenc}
\usepackage{tgpagella}

\usepackage[dvipsnames]{xcolor}
\definecolor{mycolor1}{HTML}{225CB2}
\definecolor{mycolor2}{HTML}{1E61C4}
\definecolor{mycolor3}{HTML}{113E81}
\hypersetup{
    colorlinks=true,
    linkcolor=mycolor1,
	citecolor=mycolor1,
	urlcolor=mycolor1
}

\newcolumntype{L}[1]{>{\raggedright\let\newline\\arraybackslash\hspace{0pt}}m{#1}}
\newcolumntype{C}[1]{>{\centering\let\newline\\arraybackslash\hspace{0pt}}m{#1}}
\newcolumntype{R}[1]{>{\raggedleft\let\newline\\arraybackslash\hspace{0pt}}m{#1}}

\usepackage{titlesec}

\renewcommand{\theparagraph}{\arabic{paragraph}}
\titleformat{\paragraph}[hang]{\normalsize\itshape}{\textup{\theparagraph}}{1em}{}[]

\begin{document}
\doublespacing

\begin{titlepage}
\onehalfspacing
\title{Labor Market Effects of the Venezuelan Refugee Crisis in Brazil \thanks{Hugo Sant'Anna is a Ph.D. candidate in the Department of Economics at the University of Georgia, Athens, GA, United States (hsantanna@uga.edu). Samyam Shrestha is a Ph.D. candidate in the Department of Agricultural and Applied Economics at the University of Georgia, Athens, GA, United States (samyam@uga.edu). The authors thank Carolina Caetano, Gregorio Caetano, Brantly Callaway, Roozbeh Hosseini, Nicholas Magnan, Ian Schmutte, Daniela Scur, participants at the Agricultural and Applied Economic Association (AAEA) Annual Meeting 2022 in Anaheim, CA, and participants at the Southern Economic Association (SEA) Annual Meeting 2022 in Fort Lauderdale, FL, for their comments and feedback. RAIS, the primary data used in this study is confidential and is maintained by the Federal Government of Brazil's Ministry of Labor, Employment, and Social Security, and can be requested at https://www.gov.br/pt-br/servicos/solicitar-acesso-aos-dados-identificados-rais-e-caged. The authors are willing to help acquire the data and will provide the replication code on request. Declarations of interest: None.}}
\author{Hugo Sant'Anna\thanks{Corresponding Author} \and Samyam Shrestha}
\date{\today}
\maketitle

\begin{abstract}

\noindent
\onehalfspacing
We use administrative panel data on the universe of Brazilian formal workers to investigate the labor market effects of the Venezuelan crisis in Brazil, focusing on the border state of Roraima. The results using difference-in-differences show that the monthly wages of Brazilians in Roraima increased by around 2 percent, which was mostly driven by those working in sectors and occupations with no refugee involvement. The study finds negligible job displacement for Brazilians but finds evidence of native workers moving to occupations without immigrants. We also find that immigrants in the informal market offset the substitution effects in the formal market.\vspace{0in}\\

\vspace{0in}

\bigskip
\end{abstract}
\setcounter{page}{0}
\thispagestyle{empty}
\end{titlepage}
\pagebreak \newpage

\doublespacing

\pagebreak \newpage

















\section{Introduction} \label{sec:introduction}

Forced displacement has become an escalating global crisis, with an unprecedented number of individuals compelled to flee their homes in search of safety and stability. According to the \cite{unhcr_unhcr_2023}, at the end of July 2023, there were a staggering 110 million people forcibly displaced worldwide, among whom more than 36 million were classified as refugees. Significant attention has therefore been drawn to the potential impacts of refugees on host communities. Extensive research on the labor market effects of refugees largely centers on developed nations, leaving a critical gap in understanding the effects in developing regions, where data scarcity exacerbates the challenge.

We focus on the Venezuelan humanitarian and migration crisis that started in the early 2010s. The drop in global oil prices, Venezuela's largest export, coupled with government mismanagement led to hyperinflation, a shortage of basic commodities such as food and medicine, and an upsurge in crime. As a result, more than five million Venezuelans were forced to flee the country. UNHCR estimated that by 2022, more than four million Venezuelans lived as refugees, mostly in neighboring South American countries. Brazil, in particular, had received approximately a quarter million Venezuelan refugees \citep{unhcr_venezuela_2022}, who crossed the border every year exponentially since 2013. Virtually all Venezuelans entered Brazil by land, using the only highway that connects the two countries through the Brazilian state of Roraima.

This study seeks to investigate the causal relationship between the Venezuelan refugee crisis and the labor market in Roraima, the Brazilian border state that directly experienced the influx of Venezuelans. Roraima's geographic isolation from the rest of Brazil provides a unique natural experiment. Using comprehensive administrative panel data on the universe of Brazilian formal workers and utilizing a difference-in-differences approach, we analyze the labor market effects of the refugee crisis in the Roraima labor market in comparison to similar states in north Brazil unaffected by the crisis.

Our main findings show a small but significant positive impact, of around 2 percent, on the average monthly wages of formal Brazilians who lived in Roraima. We show that this increase was not driven by an exit of low-wage Brazilian formal workers from the market as we find no significant effects on job displacement. We show, however, that this effect was primarily driven by complementary dynamics between the formal and informal sectors, and, to a lesser extent, by the fact that the presence of immigrants in the formal market allowed natives to move occupations experiencing a higher wage increase.

Nevertheless, workers in industries and occupations with a higher share of refugees did not experience any significant change in wages during the crisis, potentially due to the substitution effect of immigrants working in the same industry-occupation cells. These findings suggest that when not directly competing with native workers, Venezuelans acted as complements, increasing formal labor market wages. We also conduct analysis using nationally representative survey data and show a pronounced negative wage impact in the informal sector, more evident for individuals involved in industries with a higher immigrant involvement observed in our main administrative dataset. These results are in line with \cite{peri_task_2009, manacorda_impact_2012, dustmann_effect_2013, foged_immigrants_2016}, who argue that immigrants are imperfect substitutes for natives, have a potentially different skill set, and specialize in positive efficiency.

Refugees comprise a distinct subset of immigrants due to their more vulnerable societal position.\footnote{In the period of our study, the Venezuelans displaced by the crisis were virtually the only non-Brazilian population in the state of Roraima. Therefore, specifically for this paper, we refer to Venezuelan refugees as immigrants or foreigners, interchangeably.} Unlike other immigrant groups, their displacement is mainly involuntary.  A majority of the economic studies on the labor market impacts of refugees focus on developed countries in Europe and North America. These studies revolve around refugee shocks due to a political crisis and utilize natural experiments to analyze the labor market effects of the shock, an approach similar to ours. However, the results from these studies have little overall consensus. While some studies find no adverse effects on native employment and wages (e.g., \cite{card_impact_1990}, \cite{hunt_impact_1992}, \cite{friedberg_impact_2001}), others find large adverse effects (e.g., \cite{glitz_labor_2012}, \cite{dustmann_economics_2017}).

A small but growing subset of economic literature explores this question from the developing country perspective \citep{maystadt_winners_2014, calderon-mejia_labour_2016, ruiz_labour_2016, taylor_economic_2016, alix-garcia_refugee_2018, maystadt_development_2019}. The more recent wave of this literature revolves around the Syrian and Latin American crises. Studies on the Syrian refugee crisis have yielded varied findings that range from no effects on employment and wages \cite{fallah_impact_2019} to moderate employment losses among natives \cite{tumen_economic_2016}. Results for \cite{aksu_impact_2022} are the most similar to ours in that the study finds adverse effects on competing natives in the informal market but positive employment and wage effects for complementary workers in the formal market.

The literature on the labor market effects of Venezuelan immigrants in host countries focuses mostly on Colombia, and, to a smaller extent, on Peru and Ecuador. In Colombia, various studies reveal adverse effects on native employment within the formal labor market and reductions in wages and income in the informal sector \citep{caruso_spillover_2021, delgado-prieto_dynamics_nodate, lebow_labor_2022}.

Conversely, the labor market analysis in Peru presents a contrasting perspective. In line with our results, \cite{groeger_immigration_2024} indicates that the influx of Venezuelans to specific locations in Peru has resulted in an upswing in both employment and income among the local population. \cite{morales_venezuelan_2020} finds that an increase in the share of Venezuelan migrants in Peru is associated with an increase in the probability of being employed for Peruvians in the non-service sector although accompanied by a decrease in the probability of having an informal job.

\cite{bahar_give_2021} studies the labor market impacts of an extensive migratory amnesty program that granted work permits to nearly half a million undocumented Venezuelan migrants in Colombia in 2018. Their analysis indicates no significant impact of the program on hours worked, wages, or labor force participation of Colombian workers. Nonetheless, they study only the immediate effects of the amnesty program without considering its potential dynamic effects in the years following the treatment.

\cite{ryu_refugee_2022} addresses this question on the labor market effects of Venezuelan immigrants from the Brazilian perspective. Using a national quarterly household survey, \cite{ryu_refugee_2022} uses the synthetic control method to study the labor market impacts of the Venezuelan refugee crisis in Roraima, the affected state. The study finds that the crisis lowered labor force participation and employment rate in Roraima and did not find any effects on wages. However, the dataset did not distinguish Brazilians and Venezuelans, which can underestimate the effects of refugees.

We build upon these studies by employing a rigorous administrative panel dataset of the universe of formal workers in Brazil that has two crucial advantages over \cite{ryu_refugee_2022}. First, our data allows us to distinguish the nationality of individuals and isolate our sample to Brazilians. Second, because of its panel nature, it allows us to track individuals over time, irrespective of their mobility between states.

We organize the remainder of the paper as follows. Section \ref{sec:background} provides the background. Section \ref{sec:methods} discusses our identification strategy and empirical methodology. Section \ref{sec:data} describes the data. Section \ref{sec:results} presents the results and robustness checks of our model. Section \ref{sec:mechanism} discusses the mechanisms through which immigrants affect native wages before Section \ref{sec:conclusion} concludes.


\section{Background} \label{sec:background}
This section is divided into two parts. In the first part, we briefly overview the Venezuelan crisis. In the second, we explain the interaction between Venezuelan refugees and the Brazilian labor market, highlighting the distinctions between formality and informality.

\subsection{The Venezuelan Political and Refugee Crisis} \label{subsec:Venezuelan_crisis}

Until the early 2000s, Venezuela had one of the highest GDP per capita in Latin America. With what is considered the largest oil reserves in the world, their GDP was tied to oil exports \citep{eia_international_2019, haider_estimates_2020}. However, the drop in oil prices in the early 2010s severely hit the Venezuelan economy. Coupled with government mismanagement, it led to an unprecedented humanitarian crisis in the country with hyperinflation, a shortage of basic goods and services such as food and medicine, and a rise in crime. Although Venezuelans had already started to leave for other countries in 2011 as a result of the political crisis, the exodus skyrocketed in 2013. The United Nations High Commissioner for Refugees (UNHCR) estimated that by 2022, more than four million Venezuelans lived as refugees, mostly in neighboring South American countries. At the end of 2021, Brazil hosted 260,000 Venezuelan refugees. Due to their geographical proximity, most of the refugees in Brazil were concentrated in the state of Roraima. The Brazilian Federal Police border patrol reported that the number of Venezuelans who entered Roraima in 2017 and stayed in Brazil numbered more than 50 thousand \citep{lopes_mais_2018}. This number corresponds to 8 percent of the total population of Roraima.

In 2019, the International Migration Organization (IOM) and the Migration Policy Institute (MPI) surveyed thousands of Venezuelan refugees living in 11 Latin American and Caribbean countries \citep{echeverria-estrada_venezuelan_2020}. The report found that among the Venezuelan refugees in Brazil, almost 60 percent of the refugees were women, 55 percent were between 25-45 years, and 60 percent had secondary education. The entirety of the sampled individuals had come to Brazil by land, entering through the state of Roraima. Before migrating to Brazil, 65 percent were employed, of which 32 percent were self-employed. 26 percent were unemployed and only 5 percent were students. After immigration, only 40 percent were employed in 2019, of which 29 percent were self-employed, and 58 percent were unemployed.

Due to the only land border crossing between Venezuela and Brazil located in the Brazilian state of Roraima, the state hosted most Venezuelan refugees. Nevertheless, the insufficient public infrastructure and limited job opportunities in Roraima put a strain on the state’s health, safety, and education systems \citep{de_oliveira_use_2019}. The increasing number of refugees prompted the federal government to intervene and temporarily suspend the state's autonomy within the federation in December 2018.

The Brazilian federal government adopted a notably generous stance towards accommodating Venezuelan migrants. For instance, it established nine shelters, eight in Boa Vista. Additionally, the Brazilian government enabled Venezuelan migrants to seek employment by granting them work permits, allowing them to function as regular employees in Brazil for up to two years under temporary residency \citep{ramsey_responding_2018, ryu_refugee_2022}. Furthermore, there are policies aimed at dispersing migrants from Roraima to other states, such as Rio de Janeiro or São Paulo, starting in 2018, although this effort did not significantly impact the relocation of migrants during the periods examined in this paper.

\subsection{Brazilian Labor Market} \label{subsec:brazilian_labor_market}

Mercosul (or Mercosur in Spanish) is an economic block comprised of South American countries, including Venezuela and Brazil. Members of the block are entitled to free entry, residency rights, and the ability to work in the host country's formal labor market, subject to government authorization. Recently, the Brazilian government has offered Venezuelan refugees a special status that accelerates their permission to work in the formal labor market. Venezuelan refugees must undergo a specific process and submit certain paperwork to obtain this status, which takes at least several months.

In Brazil, any organization must have a National Legal Persons Registry number (CNPJ) to operate legally. The entity's owners must declare its purpose and intended activity to the government. If an entity is not registered, it is considered informal. The costs associated with registering a company can be relatively high, leading to a higher prevalence of informality in poorer regions. 

Legal firms are only permitted to hire workers formally, a requirement often disregarded in impoverished areas such as Roraima. Formality in the hiring sense means that employees are entitled to social security and certain rights that the employer must guarantee. Generally, workers in the formal market earn more due to these social benefits than their informal counterparts. However, employers may be at risk of labor litigation if they are caught hiring informal workers. For example, informality in northern Brazil, where we base our study, around the period of our research, corresponded to 45 percent of its total labor market \citep{azeredo_ibge_2019}. Time was likely the main cost of entry for Venezuelans to find regular employment. As more immigrants arrived and application lines increased exponentially, we posit that a fraction of these refugees procured economic activities outside formality.

\section{Methods} \label{sec:methods}
\noindent
In this section, we discuss the identification strategy and the empirical strategy used to analyze the labor market impacts of the Venezuelan refugee crisis in Roraima, Brazil. The identification strategy section highlights the geographical setting of the natural experiment, while the empirical strategy section explains the econometric models used to estimate the causal effects.

\subsection{Identification Strategy} \label{subsec:identification_strategy}

\begin{figure}[htb!]%
    \vspace{1cm}
    \centering    
    \caption{Map of Venezuela and Brazil's North Region}%
    \includegraphics[width=0.8\textwidth]{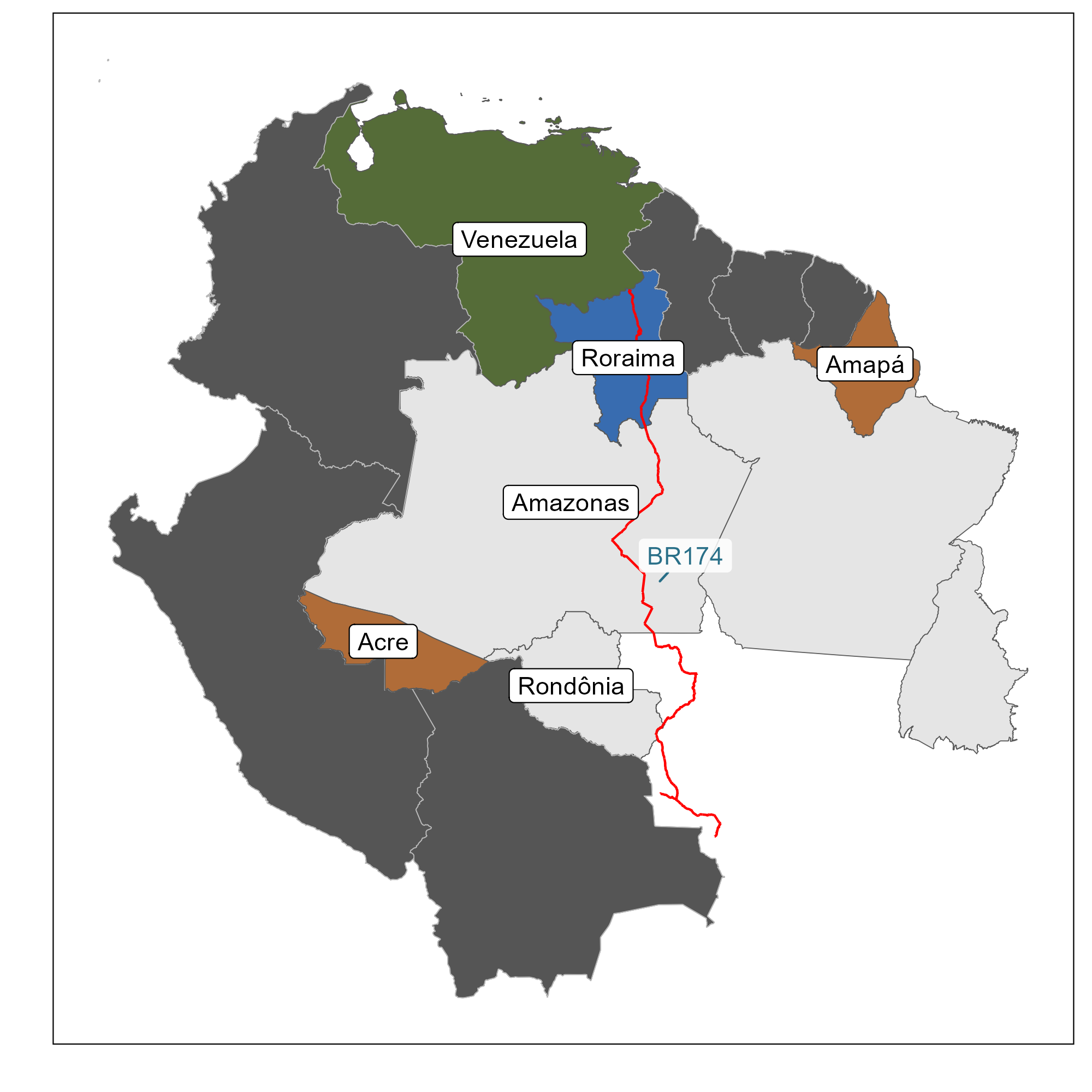}
    \label{fig:geo_overview}%
\end{figure}

Geography is crucial for our identification strategy. As shown in Figure \ref{fig:geo_overview}, Venezuela borders the Brazilian states of Amazonas and Roraima, but the Amazon rainforest makes it highly unfeasible to enter Brazil through Amazonas. Instead, people can enter Brazil from Venezuela through the state of Roraima. The BR-174 highway, represented by the solid vertical line that traverses Roraima and Amazonas in Figure \ref{fig:geo_overview}, is the only land transportation route between the two countries, which runs through Roraima from the Venezuelan border. This makes Roraima a key transit point in Brazil, but once refugees are in Roraima, they can only practically go as far as Amazonas by land. To reach larger coastal cities, they would need to use air routes. Therefore, Roraima is isolated from the rest of the northern states, which leads many refugees to choose to enter and stay there.

This setting provides a natural experiment in which Roraima acts as the treated group. For our control group, we use states that share similar sociogeographical characteristics and are located on the Brazilian border: Acre borders Peru and Bolivia, while Amapá borders Suriname and French Guiana. Like Roraima, these states also have a large portion of their population concentrated in their respective capital cities and are relatively isolated. In particular, Amapá does not have an inland connection to the rest of Brazil and can only be reached by airplane or boat crossing the Amazon hydrographic basin to reach its capital, Macapá.

According to \cite{borjas_how_1997, borjas_labor_2003, borjas_native_2006}, it may be difficult to accurately assess the impacts of immigration on the labor market by considering geography alone due to the potential for spillover effects across regions. However, in our case, Roraima is isolated with high transportation costs, which limits mobility between states or even within municipalities in the same state. 


\subsubsection{Foreign Presence in RAIS}\label{subsubsec:foreign_presence}

\begin{figure}[htb!]
\centering 

\begin{subfigure}{\linewidth}
    \centering 
    \includegraphics[width=0.8\textwidth]{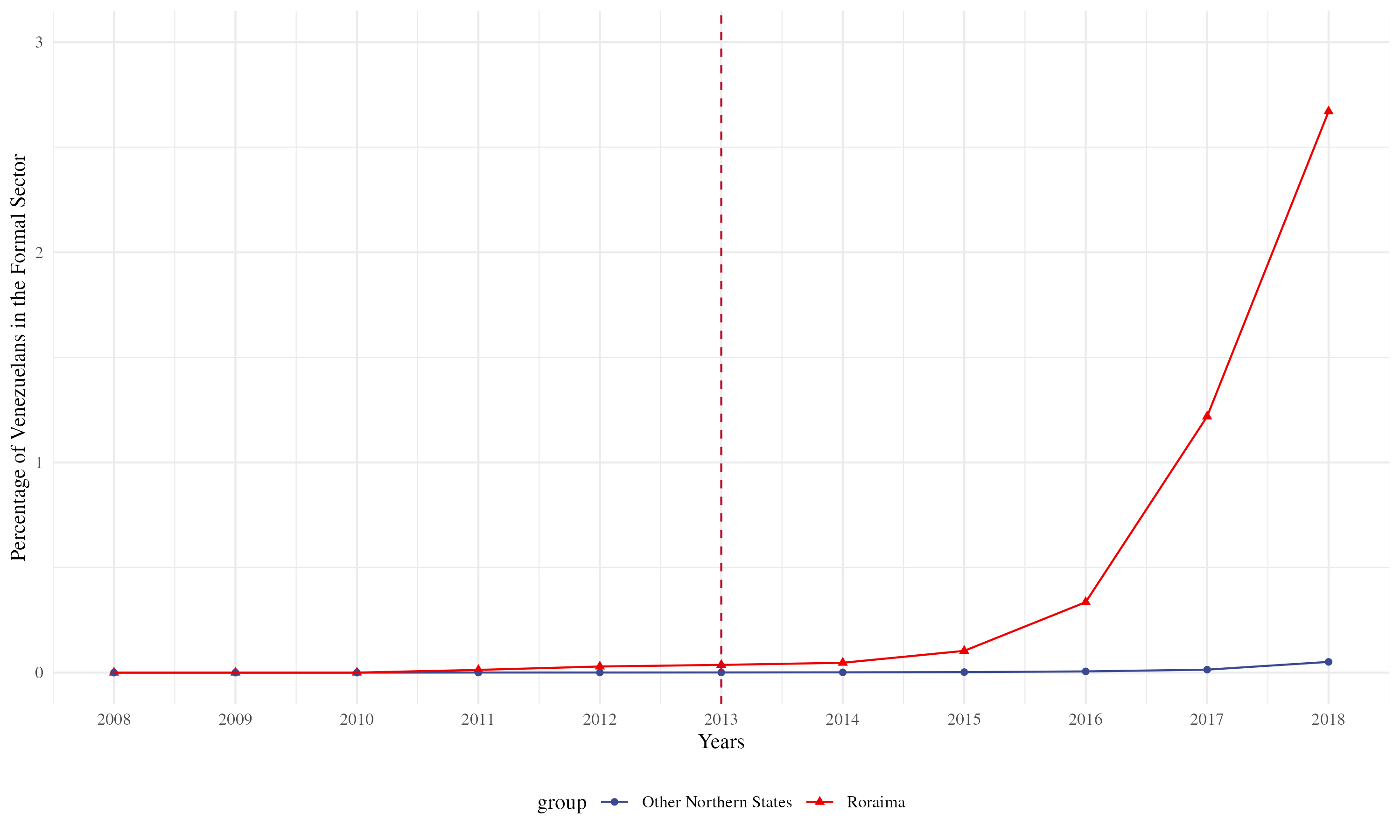}
    \caption{Venezuelans as \% of Labor Market}
    \label{fig:venezuelantrend}
\end{subfigure}%

\vspace{0.5cm} 

\begin{subfigure}{\linewidth}
    \centering 
    \includegraphics[width=0.8\textwidth]{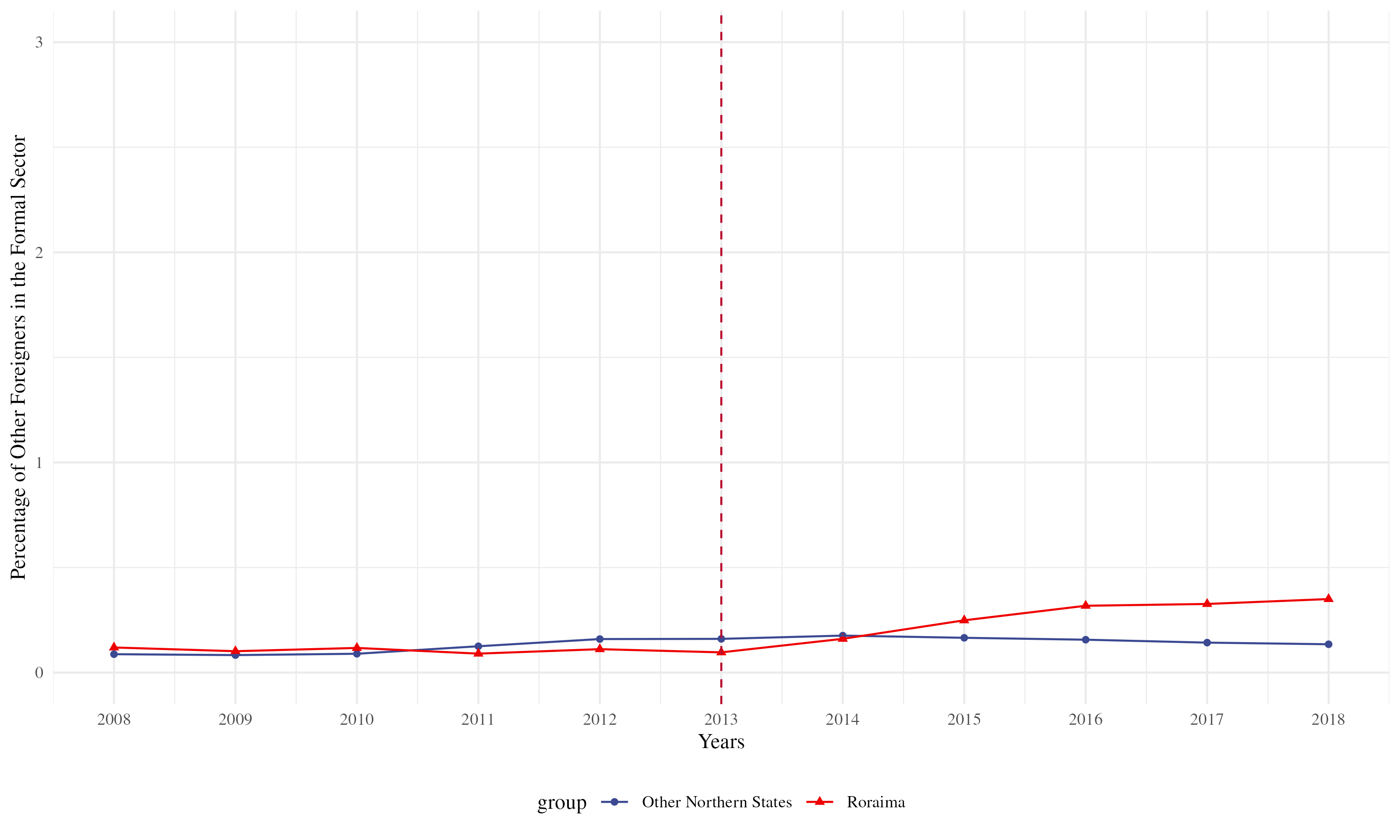}
    \caption{Other Foreigners as \% of Labor Market}
    \label{fig:otherforeignerstrend}
\end{subfigure}%

\caption{Proportion of non-Brazilians in the formal labor market for Roraima and the control states}
\label{fig:foreigntrends}
\end{figure}

A key assumption of our paper is that refugees were not attracted to high-growth regions, but ended up in Roraima due to its geographical isolation and high transportation costs to move to cities farther from the border. To verify, we can use our administrative data to check the location of Venezuelan refugees in our sample. We can also examine wage trends before the crisis to confirm that Roraima’s growth trend was not significantly different from the control regions, which we explore in the next subsection. Figure \ref{fig:venezuelantrend} shows the annual percentage of Venezuelans in the labor market, grouped by treatment and control states. In 2017, Venezuelans represented about 2 percent of Roraima’s formal labor market, while none were observed in Amapá or Acre. Figure \ref{fig:venezuelantrend} supports our assumption by demonstrating that Venezuelans largely remained in Roraima.

It would be problematic for our experiment setting if any other nationality besides Venezuelans had a substantial growth in population, either in Roraima or the control states. To ensure that the exponential growth of foreigners we observe is due to Venezuelan refugees, we construct Figure \ref{fig:otherforeignerstrend} by aggregating data on all immigrant nationalities other than Venezuelans and plotting their proportion in the formal labor market. As the figure shows, there is negligible growth in the control states and no significant growth in Roraima following 2014. The non-zero values we observe for non-Venezuelan immigrant nationalities can be attributed to several factors. First, there is a significant presence of Haitian immigrants in control states due to United Nations peace operations in Haiti that began in the 2000s and remained constant over the study years. Second, there are observations of other Latin American and Bolivian nationals in the data in control states due to the proximity of Acre to Bolivia and the resulting natural population exchange. Finally, some of these foreigners may be Venezuelans with dual citizenship, as they only appeared in Roraima after the crisis began, as Figure \ref{fig:otherforeignerstrend} shows. Since there is marginal change in these populations in either treatment or control groups, the non-zero values should not pose a threat to our identification strategy.

Although the administrative data provides a good overview of Venezuelans entering Brazil and remaining in the state of Roraima, it only includes those in the formal labor market. It could compromise our identification strategy if the control states also had many Venezuelan refugees outside the formal labor market. In Appendix \ref{sec:vz_permanence_in_rr}, we use refugee application data to demonstrate that this is not the case. We show that only Roraima, not the control states, experienced a significant increase in the number of Venezuelan citizens seeking refugee status applications.

\subsection{Empirical Strategy}\label{subsec:empirical_strategy}

To empirically assess the effects of the refugee crisis in Roraima labor market, we estimate equation (\ref{eq:wagesdid_bin}).
\begin{equation} \label{eq:wagesdid_bin}
y_{imt} = \theta_{i} + \alpha_t + \beta^{bin} D_{imt} + \epsilon_{imt}
\end{equation}

In this model, $y_{imt}$ denotes the logarithm of the average monthly wage for individual $i$ working in municipality $m$ during year $t$. The term $\theta_i$ captures time-invariant characteristics unique to each worker, which in our dataset are represented by a series of social identifier dummies. The $\alpha_t$ term represents year fixed effects, accounting for broader temporal trends and seasonality that could affect wages across all states.

Our primary focus is on the parameter $\beta^{bin}$, which quantifies the returns of being exposed to Venezuelan immigration after 2013. This exposure is encapsulated in the binary treatment variable $D_{imt}$, which takes a value of 1 for individuals in the municipality that belongs to Roraima post-2013 and zero otherwise. The term $\epsilon_{imt}$ denotes the idiosyncratic error term, capturing unobserved influences on wages. Since our observed treatment heterogeneity can occur across regions within each state, we cluster at the municipality level, as discussed in \cite{abadie_when_2023}.

To further refine our analysis, we introduce an alternative specification where the treatment effect is modeled as a continuous variable, following \cite{callaway_difference--differences_2021}. This approach uses the ratio of Venezuelan refugees per municipality as a proxy for exposure to the immigration shift, as stated in equation (\ref{eq:wagesdid_cont}). Here, $R_{imt}$ represents the ratio of immigrants per municipality $m$ where individual $i$ is located at a given year $t$.
\begin{equation} \label{eq:wagesdid_cont}
y_{ist} = \theta_{i} + \alpha_t + \beta^{cont} R_{imt} + \epsilon_{ist}
\end{equation}

Both approaches exploit the within-individual variation in the exposure to whether the state experienced the refugee crisis. Our identifying assumption is that the time-varying shocks are orthogonal to the treated state.

\subsubsection{Doubly-Robust Estimator}

To strengthen the credibility of our causal inference, we employ a doubly robust estimation framework, balancing observable characteristics of workers in our data through propensity scores and applying the inverse of these scores as weights in our regressions. This approach ensures that particular features across control and treatment groups will not bias our results. Moreover, it has the advantage that our results are consistently estimated even if one of the difference-in-differences or the propensity score specifications is not \citep{uysal_doubly_2015, santanna_doubly_2020}.

In the first step of the doubly robust framework, we use the following logistic regression:
\begin{equation}
p(X_{i}) = Pr(Z_i = 1 | X_i) = F(\theta'X_{i}) = \frac{e^{\theta'X_{i}}}{1 + e^{\theta'X_{i}}}
\end{equation}

In this regression, $P(Z_i = 1 | X_i)$ represents the probability of being in the treatment group given the vector $X_i$ of time-invariant covariates for individual $i$ and $Z_i$ represents treatment group membership, which is working in Roraima. The parameter $\theta$ establishes the logistic relationship between the covariates and the group membership.

The second step is to construct the inverted propensity scores based on control and treatment group.
\begin{equation} \label{eq:weights}
    w_i = 
    \begin{cases} 
        \frac{1}{1 - p(X_i)}, & \text{if } Z_i = 0 \\
        \frac{1}{p(X_i)}, & \text{if } Z_i = 1
    \end{cases}
\end{equation}

We also impose trimming rules to avoid propensity score values that are too extreme, avoiding weights above the 99.75 percent quantile \citep{lechner_practical_2019}. The final step minimizes the weighted sum of squares in the two-way fixed effects framework. For instance, in our binary treatment case, we have:
\begin{equation} \label{eq:drdid}
\hat{\beta}^{dr}_{bin} = \argmin_{\beta} \sum_{i = 1}^N w_{i} (y_{imt} - \beta D_{imt} - \gamma_{i} - \omega_{t})^2
\end{equation}

Where $w_{i}$ is calculated using Equation (\ref{eq:weights}). $N$ is the number of observations in the sample. The analogous approach is employed for the continuous treatment, in which we replace $D_{imt}$ to $R_{imt}$.


\section{Data} \label{sec:data}

Our primary dataset is the Annual Social Information Survey (RAIS) maintained by the Brazilian Ministry of Labor and Employment. RAIS contains the universe of the Brazilian formal labor market, comprising panel information on individual workers and establishments, including employment status, occupation, industry\footnote{Throughout this paper, we use the term `economic sector', `industry', or 'economic activity' to refer to the industry in which a firm operates, such as retail, construction, or restaurants. We use the term `occupation' to describe the specific job title or role of the worker within that sector, such as janitor, accountant, or waiter.}, and wages. We focus on observations from 2008 to 2018. Our pre-treatment years are 2008-13, and the post-treatment period includes 2014-18. We use this cutoff since the refugee crisis intensified in 2014 based on the data and journalistic accounts (see the identification strategy section for more details).

The Brazilian equivalent of a Social Security Number, PIS (Social Integration Program), identifies unique persons and is present in the data. Along with information on the labor market at the individual level, the data also provides sociodemographic information, including nationality. For occupation, we use the first three digits of the Brazilian Occupation Code (CBO) to categorize, while for the economic sector, we use the first 2 digits of the National Registry of Economic Activities (CNAE).

The dataset is constructed based on the contract generated between the employee and the firm. Therefore, the base observation is contract-year. To convert the cells to individual-year observations, we restricted the data to Brazilians in a given year employed in the private sector with more than or equal to 30 weekly hours of labor. If there are multiple observations for the same individual, we use the contract with the oldest tenure and the highest pay \citep{lavetti_gender_2023}.

In our dataset, payments are recorded as the average monthly compensation each individual receives according to their contract. To construct our primary outcome variable, we exclude individuals with recorded payments of zero. Furthermore, we apply censoring to this variable at both tails of the distribution: at the 99.75th percentile from the upper end and the 0.25th percentile from the lower end. This measure mitigates the influence of extreme outliers. Finally, we transform the variable into an hourly wage measure by integrating it with the recorded weekly hours worked times four.

\subsection{Summary Statistics}

By restricting our sample to Brazilian workers who were employed in the private sector with at least 30 weekly hours of labor, we retain the sample size of 58,379 individuals, of whom 12,581 are from Roraima and 45,798 from control states. Table \ref{Tab:summary_capital} summarizes the socioeconomic and demographic characteristics of Brazilians in our treatment state, Roraima, and our control states in the RAIS datasets in 2013, the year right at the start of the refugee crisis. The average monthly wages for Roraima and control states were comparable at R\$ 1,916 and R\$ 1,841 respectively. These wages are reported in local currency adjusted to 2018 R\$. In 2018, the national minimum wage was around R\$ 1,000. The average earnings in these states were, on average, twice this value. Other variables such as average age, experience, fraction of female workers, racial composition, and workers by economic sector and occuption type  were all comparable between Roraima and the control states.

\begin{table}[htb!]
\vspace{1cm}
  \centering
 \caption{Descriptive Statistics of Natives from Roraima and Control States}
    \label{Tab:summary_capital}

\begin{threeparttable}

\begin{tabular}[t]{lrr}
\toprule
& (1) & (2) \\
& Roraima & Control States\\
\midrule
Mean monthly wage  & \num{1916.07} & \num{1841.05}\\
Mean age & \num{36.92} & \num{37.85}\\
Mean experience & \num{60.75} & \num{63.70}\\
\addlinespace
Fraction of females & \num{0.36} & \num{0.32}\\

\addlinespace
Race by fraction: & & \\
\ White  & \num{0.22} & \num{0.17}\\
\ Black  & \num{0.02} & \num{0.03}\\
\ Mixed  & \num{0.63} & \num{0.65}\\
\ Color not declared & \num{0.12} & \num{0.15}\\

\addlinespace
Education by fraction: & & \\
\ High school dropouts  & \num{0.24} & \num{0.32}\\
\ High school graduates  & \num{0.64} & \num{0.58}\\
\ College graduates  & \num{0.12} & \num{0.10}\\

\addlinespace
Economic activities by fraction: & & \\
\ Commerce  & \num{0.41} & \num{0.38}\\
\ Construction  & \num{0.08} & \num{0.09}\\
\ Extraction industries  & \num{0.03} & \num{0.05}\\
\ Hotels and restaurants  & \num{0.04} & \num{0.03}\\
\ Manufacturing and utilities  & \num{0.10} & \num{0.10}\\
\ Other services  & \num{0.35} & \num{0.35}\\
\addlinespace
Job occupations by fraction: & & \\
\ High school level technicians  & \num{0.07} & \num{0.08}\\
\ Factory occupations  & \num{0.22} & \num{0.27}\\
\ Services  & \num{0.43} & \num{0.41}\\
\ Retail and wholesale  & \num{0.14} & \num{0.11}\\
\ Rural occupations  & \num{0.02} & \num{0.03}\\
\ Scientific and liberal arts  & \num{0.07} & \num{0.05}\\
 \addlinespace
  N & \num{12581} & \num{45798}\\
\bottomrule
\end{tabular}
    \begin{tablenotes}
        \begin{footnotesize}
        \item \textit{Note:} Observations represent full-time workers working in 2013 for both Roraima and the Control States (Amapá and Acre). Mean experience in months. Economic activities are classified by the first two digits from their respective  code. Occupations are likewise classified by the first three digits.
        \end{footnotesize}
    \end{tablenotes}
\end{threeparttable}
\end{table}

Figures \ref{fig:industries_vz} and \ref{fig:occupations_vz} illustrate Venezuelan formal workers by economic sectors and occupations respectively in Roraima using 2018 RAIS data, the final year in our data and the year with the largest number of Venezuelans observed in the data. Figure \ref{fig:industries_vz} shows that most of the Venezuelans in Roraima were in the retail commerce sector, followed by restaurants, construction, wholesale commerce, and gardening and landscaping. Figure \ref{fig:occupations_vz} shows that most of them worked as general service employees, salespersons, machine operators, and office clerks, which are semi-skilled or low-skilled.

Table \ref{tab:sumstatvz2018} compares Brazilians with Venezuelans in the formal labor market in Roraima in 2018, the year with the most Venezuelans in our sample period. While the reported race is comparable between the two groups, a much higher percentage of Venezuelans reported to be of mixed race. While men and women were almost equally represented among Brazilians in the Roraima formal labor market, it was much more dominated by men for Venezuelans, at around 76 percent. This comes in stark contrast to the fact that around 60 percent of the Venezuelan refugee population in Roraima was female \citep{echeverria-estrada_venezuelan_2020}, highlighting that male Venezuelans were more active in the formal labor market.

Venezuelans tend to have a higher proportion of individuals with only high school education and a markedly lower percentage of college graduates compared to their Brazilian counterparts. Table \ref{tab:sumstatvz2018} also highlights the top five economic activities and occupations with the highest Venezuelan representation. This was determined by calculating the ratio of Venezuelans to total workers in each category, using the top 2 digits of both to allow for a broader categorization. Pictures \ref{fig:industries_vz} and \ref{fig:occupations_vz} were generated using the same procedure.

The educational profiles of the two groups are mirrored in the sectors they predominantly engage in. Immigrants are primarily employed in industries like commerce, restaurants, and construction, which typically require less specialized education. They are mostly found in industries such as restaurants, retail, and construction, encompassing auxiliary jobs across these various industries through the ``General Service'' category in occupation. In contrast, Brazilians exhibit a more diverse industrial distribution, notably in sectors we categorize as ``Other Industries'', which includes professions in teaching, banking, legal services, healthcare, and more. Additionally, a significant portion of Brazilians occupy mid-level office positions, classified as 'Office Clerks'.

Finally, the data also reveal a stark contrast in earnings between natives and immigrants. On average, Venezuelans are younger and earn significantly lower wages than Brazilians. This wage disparity may reflect differences in job types, experience levels, or the indirect costs of immigration.

\begin{table}[htb!]
  \centering
  \vspace{-0.2in} 
  \begin{threeparttable}
    \caption{Statistics of Natives and Venezuelans in 2018}
    \label{tab:sumstatvz2018}
    \begin{tabular}{lcc}
      \toprule
      & (1) & (2) \\
      Percentage of & Brazilian & Venezuelan \\
      \midrule
      Race & & \\
      \addlinespace
      White  & 6.18 & 5.85 \\
      Black  & 1.07 & 1.76 \\
      Indigenous  & 0.33 & 0.03 \\
      Mixed  & 34.04 & 52.94 \\
      Not Declared  & 58.38 & 39.43 \\
      \midrule
      Gender & & \\ 
      \addlinespace
      Female  & 48.68 & 23.96 \\
      Male  & 51.32 & 76.04 \\
      \midrule
      Education & & \\
      \addlinespace
      No High School  & 16.97 & 16.29 \\
      High School  & 54.97 & 76.02 \\
      College  & 28.05 & 7.69 \\
      \midrule
      Industry by Economic Activities\tnote{1} & & \\
      \addlinespace
      Retail Commerce  & 16.98 & 38.01 \\
      Restaurants & 2.74 & 15.90 \\
      Construction & 2.56 & 6.59 \\
      Wholesale Commerce & 3.11 & 5.68 \\
      Gardening and Landscaping & 3.59 & 1.70 \\
      Other Industries & 71.03 & 32.13 \\
      \midrule
      Occupations\tnote{1} & & \\
      \addlinespace
      General Service Employees & 17.37 & 34.69 \\
      Retail and Wholesale Salesperson & 8.65 & 14.42 \\
      General Construction Workers & 2.93 & 10.76 \\
      Machine Operators &  4.23 & 6.19 \\
      Office Clerks & 28.80 & 5.59 \\
      Other Occupations & 38.03 & 28.36 \\
      \midrule
      Mean Age & 36.86 & 31.67 \\
      Mean Wage & 2857.94 & 1198.41 \\
      Total Observations & \num{128389} & \num{3523} \\
      \bottomrule
    \end{tabular}
    \begin{tablenotes}
      \footnotesize 
      \item [1] See text. Occupation and industry identifiers from the data are provided by the Registry of Brazilian Occupations and Registry of Economic Activities. For both variables, we use the code top 2 digits to generate the table. However, for our regressions, we allow occupations to be detailed by their top 3 digits. 
    \end{tablenotes}
  \end{threeparttable}
\end{table}

\section{Results} \label{sec:results}

In this section, we present the main results of our analysis. In the first set of regressions, we explore the effects of Venezuelan immigration on the logarithmic average monthly wage of natives using a difference-in-differences strategy.

\subsection{Main Results}

Columns (1) and (2) in Table \ref{tab:mainresults} report our coefficients of interest, $\beta^{bin}$ and $\beta^{cont}$ respectively, representing the returns on wages by being in Roraima after the Venezuelan crisis. On average, Roraima experienced a slight positive wage increase after the influx. The interpretation of the continuous approach is that for every 1 percent increase in the presence of Venezuelans in the formal labor market, the wage increased by 2 percent, also consistent with the 2.2 percent increase in the binary treatment variable regression.
\vspace{0.3in}
\begin{table}[htb!] \label{tab:mainresults}
\centering
\caption{Main Results - Roraima x Amapá and Acre}
\label{tab:mainresults}
\begin{threeparttable}

\begin{tabular}[t]{lcccc}
\toprule
\multicolumn{1}{c}{ } & \multicolumn{2}{c}{Log Wage} & \multicolumn{2}{c}{Job Retainment} \\
\cmidrule(l{3pt}r{3pt}){2-3} \cmidrule(l{3pt}r{3pt}){4-5}
  & (1) & (2) & (3) & (4)\\
\midrule
Treat: Binary & \num{0.022}*** &  & \num{-0.010} & \\
 & (\num{0.006}) &  & (\num{0.010}) & \\
Treat: VZ Ratio x 100 &  & \num{0.021}*** &  & \num{0.001}\\
 &  & (\num{0.003}) &  & (\num{0.005})\\
\midrule
Individual FE & X & X & X & X\\
Year FE & X & X & X & X\\
\midrule
R2 Adj. & \num{0.875} & \num{0.875} & \num{0.150} & \num{0.150}\\
RMSE & \num{0.245} & \num{0.245} & \num{0.370} & \num{0.370}\\
N Clusters & \num{53} & \num{53} & \num{53} & \num{53}\\
N & \num{577552} & \num{577552} & \num{577552} & \num{577552}\\
\bottomrule
\end{tabular}

\begin{tablenotes}
\footnotesize 
\item [1] Standard-errors are clustered by municipality. 
\item [2] Propensity score explanatory variables are gender and race indicators, age, age-squared, tenure, tenure-squared and education level. 
\item [3] * $p < 0.1$, ** $p < 0.05$, *** $p < 0.01$
\end{tablenotes}
\end{threeparttable}

\end{table}

To address the possibility that our observed wage effects are influenced by low-wage earners exiting the labor market due to immigrant replacement, we employ a linear probability model in line with equations (\ref{eq:wagesdid_bin}) and (\ref{eq:wagesdid_cont}). Our focus is to evaluate the likelihood of a worker departing from their job within a given year, a key concern in the context of immigrant labor dynamics. This aspect of our analysis helps determine whether wage increases could be falsely appearing as a result of lower-paid workers being replaced by immigrants. We identify job departures in our dataset through a specific dummy variable, which indicates whether a worker remained with the same firm until the end of the year.

The results are detailed in columns (3) and (4) of Table \ref{tab:mainresults}. Notably, we cannot statistically reject the null hypothesis of the immigrant crisis affecting job retention in Roraima. These findings provide strong evidence that the observed positive wage effects are not merely a consequence of lower-wage earners exiting the labor market post-immigration. This stability is the first step to indicate that the wage increase is not artificially inflated by a reduction in the number of low-wage positions but rather reflects a genuine enhancement in wage levels. Such a scenario aligns with economic theories that posit immigration can have a complementary effect on the native workforce, potentially due to increased labor demand and productivity. Moreover, it could also be the complementarity effect of informal and formal sectors due to a larger concentration of immigrants outside the formal market. We will discuss these mechanisms in more detail in the following sections.

\cite{ryu_refugee_2022} finds that while labor force participation decreased among working-age individuals in Roraima, there were no significant impacts on hourly wages in Roraima. Their use of the cross-sectional PNAD-C data drives the differences in the results as PNAD-C cannot distinguish the nationality of individuals in the sample, therefore including recently arrived Venezuelan migrants in the analysis, who were either searching for jobs or were involved in transitory low-wage jobs. This inherently overestimates the impacts on labor force participation and underestimates the impacts on wages. Our results come from the longitudinal individual-level data distinguishing Brazilians and non-Brazilians, eliminating this issue, and ultimately illustrating positive wage effects in Roraima and the absence of job loss among formal workers.

The key assumption of our strategy is that in the absence of the treatment, our outcome variable will exhibit identical trends in treatment and control states. While we cannot test the counterfactual in post-treatment years, the pre-treatment trends between the treated and control for the outcome variable should be parallel over time.

To test this assumption of parallel trends, we employ an event study method using the binary treatment variable where $\beta^{bin}$ is disaggregated for every year present in the data. The reference year for comparison is 2013, the year before the Venezuelan crisis and the refugee influx gained momentum. The estimation procedure is shown in Equation (\ref{eq:wageses}), where the indicator function $D_{imt}$ now takes a separate value for each year within the summation. $\beta_{t}^{dr}$ explains the average difference between treatment and control groups for each year $t$ compared to the reference year of 2013.
 \begin{equation} \label{eq:wageses}
    y_{imt} = \sum^{2018}_{\substack{t = 2008 \\ t \neq 2013}} \beta^{dr}_{t} D_{imt} + \theta_i + \alpha_t + \epsilon_{imt}
\end{equation}

\begin{figure}[htb!]
    \centering
        \includegraphics[width=0.8\textwidth]{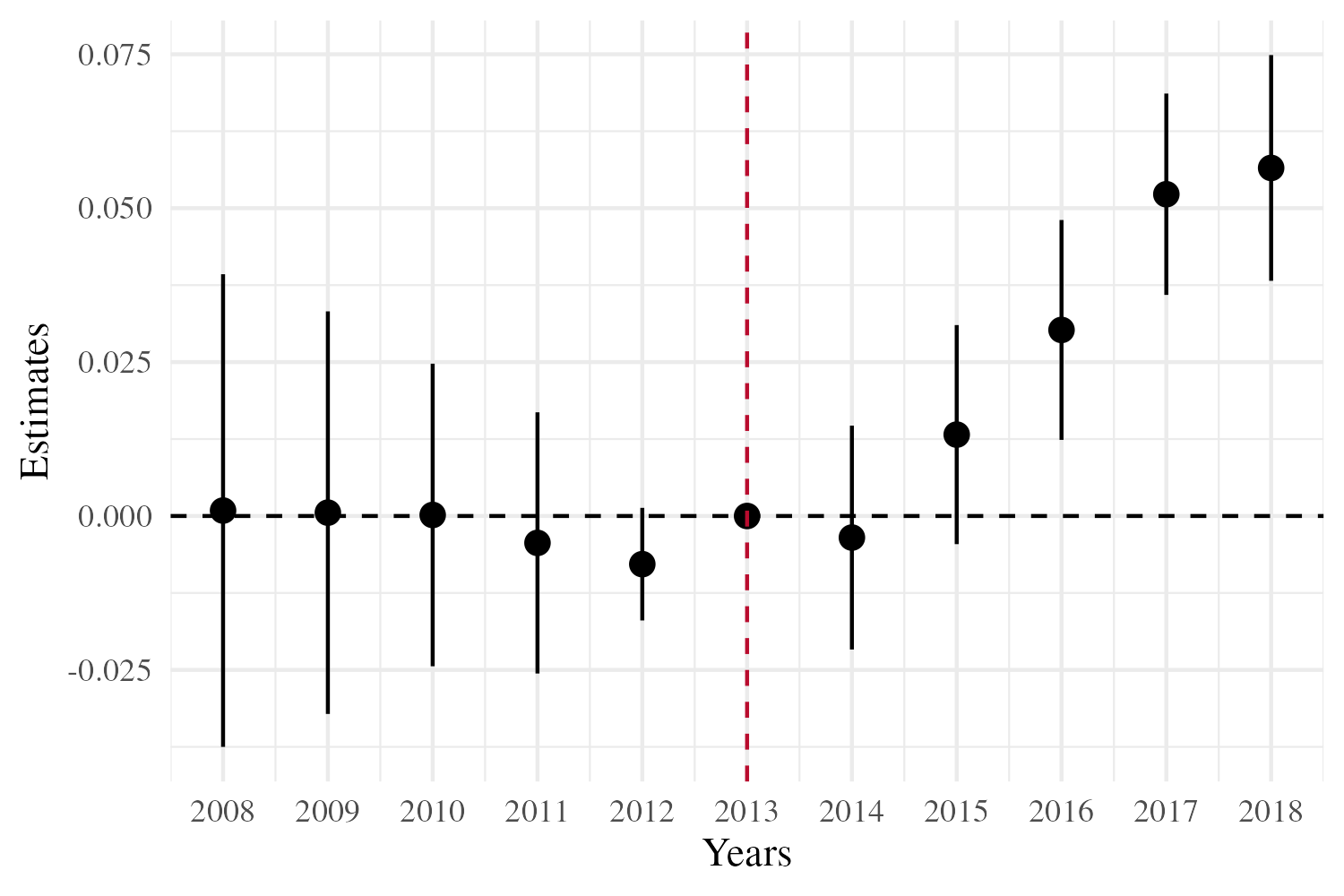}  
        \caption{Log Monthly Wages Effects Event Study }
        \label{fig:mainresults_es}
\end{figure}

If control and treated groups are comparable before treatment, then, on average, $\beta_{t}=0$ for $t \in \{2008, \ldots, 2012\}$. In other words, there should be no difference in trends between treatment and control groups before the treatment. Assuming the only disturbance in Roraima's job market after 2013 is the immigration flow, any variation in the post-treatment period estimates in Roraima must be associated with the refugee crisis. If the refugee crisis affects wages, then the effects should be increasing over time due to the increase in the total refugee numbers. Figure \ref{fig:mainresults_es} shows the event study estimates, which precisely illustrate this pattern.

The results reveal a parallel trend in earlier years right before treatment. They yield pre-treatment estimates not statistically different from zero, with an increasingly upward trend in post-treatment periods. This suggests the difference-in-differences model captured the positive effect of Venezuelans in the formal market, allowing Brazilians, on aggregate, to increase their wages.

\subsection{Synthetic Control Method}

Concerns may arise in our estimation strategy of using the doubly robust difference-in-differences method given the nature of our natural experiment with only one treatment state (Roraima) if the control group is chosen incorrectly. In the main analysis, we compare the treatment state with carefully chosen control states in Brazil with states with socioeconomic and sociodemographic characteristics comparable to those of the treatment state. We also showed pre-treatment parallel trends in our outcome variable. However, it may lead to biased estimates if the control groups are characteristically different from the treatment state.

A popular way of estimating the causal effects of the treatment in the case of a singular or a few treatment states is by using the Synthetic Control Method (SCM). Using SCM with the northern Brazilian states as our donor pool for the synthetic Roraima, we confirm the validity of our choice of control states in the main analysis. We present the details in Section \ref{sec:scm_appendix}, but from Fig \ref{fig:fig_scm_lw}, we observe perfect parallel trends between Roraima and synthetic Roraima until 2013 and a few years following. The trends started to diverge in 2016 where the wages for synthetic Roraima have a reduction in slope but Roraima's slope reduction is less steep. In the year 2018, we see an almost 0.20 percentage point difference between Roraima and synthetic Roraima. This exercise confirms the results from our primary analysis using difference-in-differences. We also provide results from the Synthetic Difference-in-Differences model (SDID) \citep{arkhangelsky_synthetic_2021} in Section \ref{sec:sdid} where the outcomes closely mirror our primary results showing an upward wage trend in Roraima compared to synthetic Roraima.

\section{Mechanisms} \label{sec:mechanism}

In this section, we explore the mechanisms through which Venezuelan immigration might be elevating overall wages in Roraima's formal labor market. We identify three primary mechanisms through which this impact could manifest:

\begin{enumerate}
    \item Immigrants in the formal labor market can affect wages by increasing worker efficiency and offsetting any substitution effects, resulting in an overall positive change.
    \item The informal sector, presumably with a larger concentration of immigrants, can increase overall wages in the formal sector through complementarity.
    \item The presence of the immigrant population as a whole, which by 2018 corresponded to 10 percent of Roraima's population, can increase native wages by boosting consumption and ultimately labor demand. However, we cannot currently test for this with our labor market data.
\end{enumerate}

\subsection{Effects Due To Presence of Immigrants in the Formal Sector}

Economists often consider that immigration is not evenly balanced across groups of workers. For example, if high school graduates are the majority of Venezuelan immigrants, they potentially compete with native high school graduates, but not necessarily with individuals holding a college degree \citep{card_is_2005, card_immigration_2009, borjas_wage_2017, llull_immigration_2018}. Another dimension is occupation, wherein immigrants tend to occupy manual-intensive or low-skilled jobs \citep{foged_immigrants_2016}, which could increase the efficiency of the market and allow the overall average wages to grow. We test for treatment heterogeneity by education and by industry and occupation with immigrant presence.

\subsubsection{Treatment Heterogeneity by Education}

We now explore potential treatment heterogeneity by education. We categorize our data into three groups: individuals with a college education, individuals with a high school education but no higher degree, and individuals with less than a high school education. Given that most of the observed Venezuelan immigrants in RAIS are high school graduates, we anticipate distinct effects within this cohort.

\begin{figure}[ht]
    \centering
    \begin{subfigure}[b]{0.48\textwidth}
        \centering
        \includegraphics[width=\textwidth]{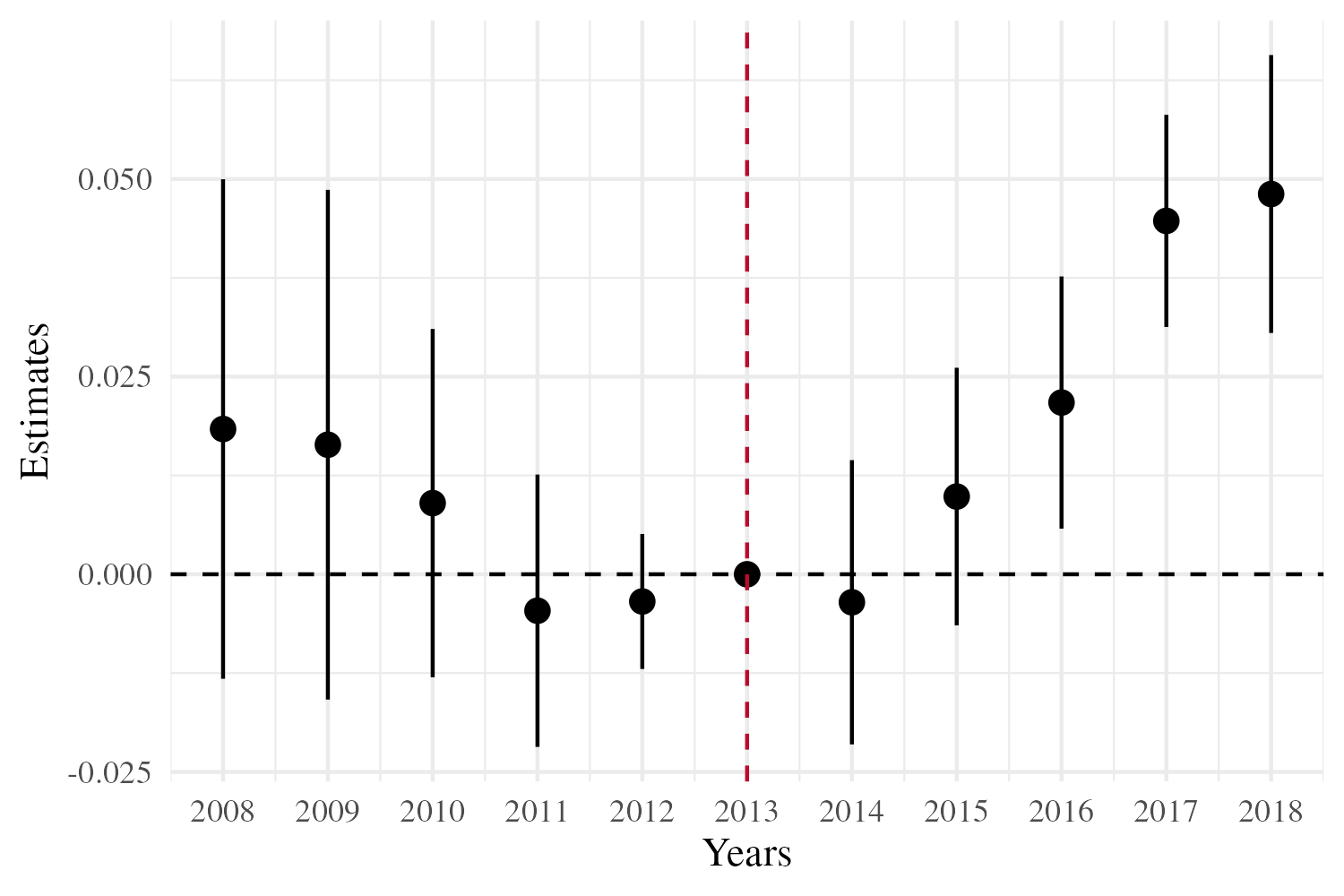}
        \caption{High School}
        \label{panel:A}
    \end{subfigure}
    \hfill
    \begin{subfigure}[b]{0.48\textwidth}
        \centering
        \includegraphics[width=\textwidth]{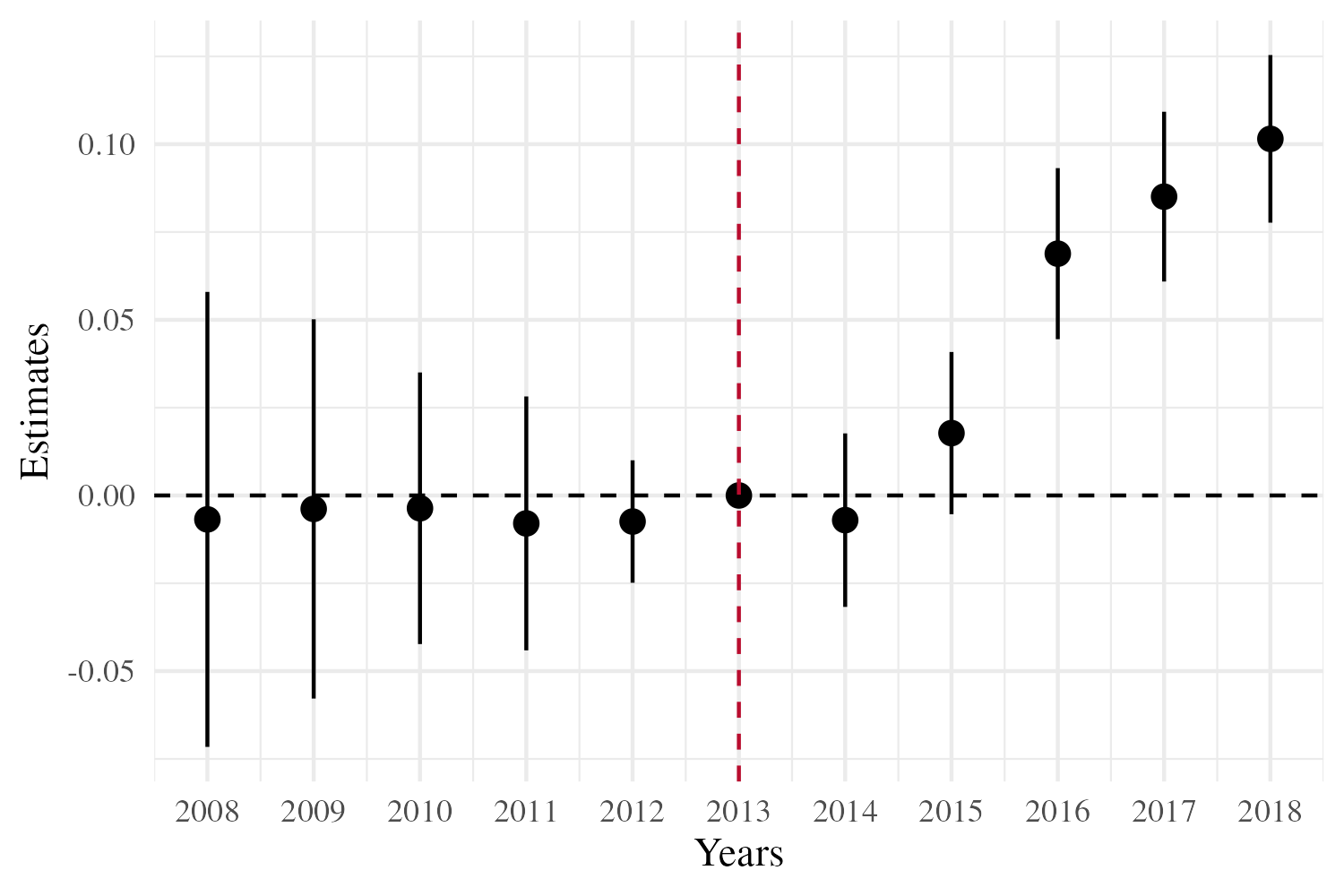}
        \caption{Low Education}
        \label{panel:B}
    \end{subfigure}

    \begin{subfigure}[b]{0.6\textwidth}
        \centering
        \includegraphics[width=\textwidth]{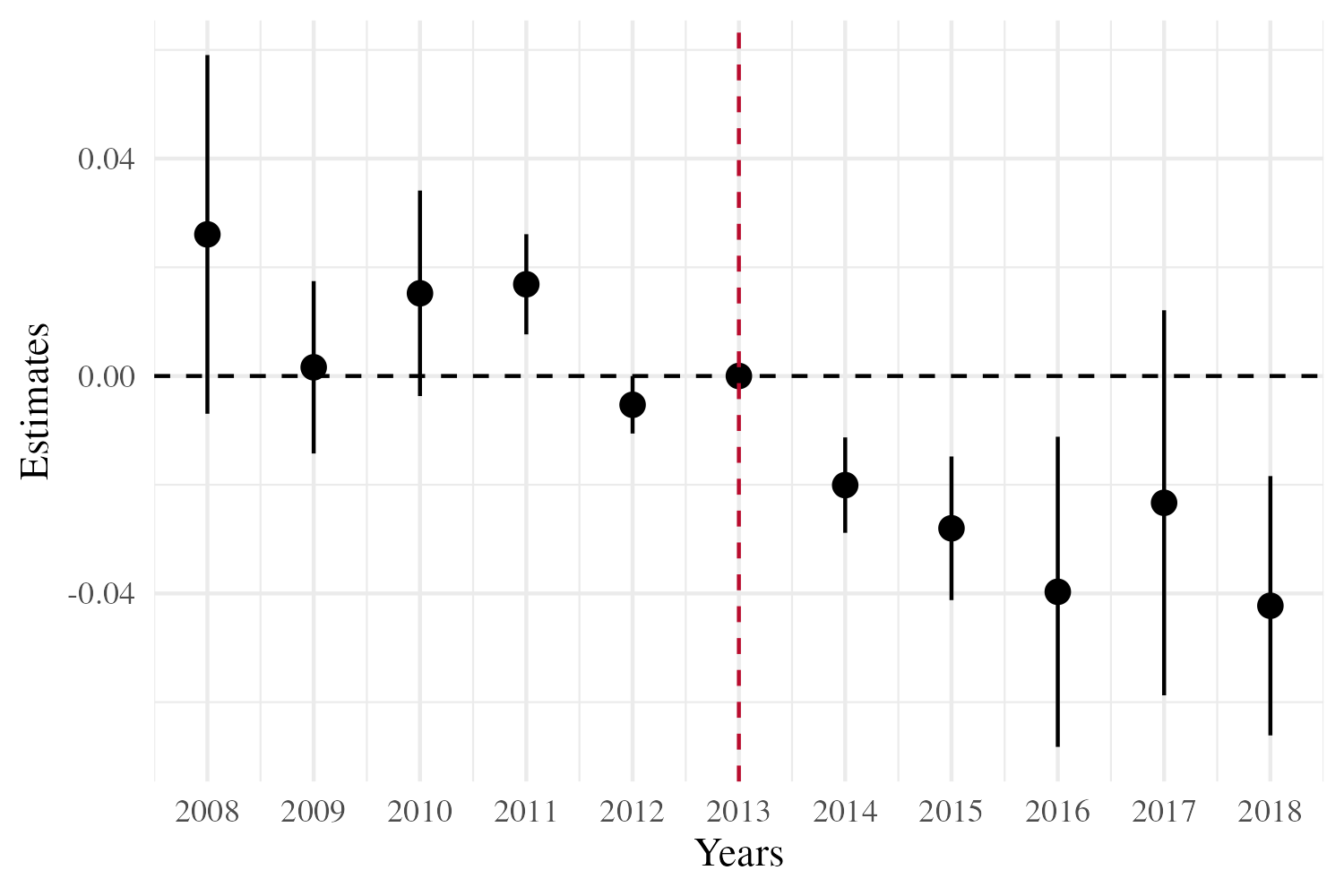}
        \caption{College}
        \label{panel:C}
    \end{subfigure}

    \caption{Event Studies for Heterogeneous Treatment Effects by Education Cohorts}
    \label{fig:education_cohorts}
\end{figure}

Table \ref{tab:education} presents the results of our difference-in-differences analysis, considering the heterogeneous effects by education level. Concurrently, the corresponding event studies are depicted in Figure \ref{fig:education_cohorts}. The two rows of the table delineate the results using the binary treatment variable and the continuous variable respectively. Columns (1) and (2) use a sample of Brazilians with at least a college education. Results indicate approximately a 3.7 percent wage decrease under the binary treatment variable and a 1.3 percent decrease per 1 percent increase in the presence of Venezuelans in Roraima. While college-educated individuals in Roraima appear to be the most adversely affected by immigration, the downward trend in college graduate wages, as illustrated in Figure \ref{panel:C}, suggests that this negative impact is compounded by pre-immigration variations.

Columns (3) and (4) use a sample of individuals with a high school education or more. Results show a wage increase of around 1.5-1.7 percent for both types of treatment variables. These figures are markedly lower in magnitude compared to their counterparts with less than high school education, as shown in columns (5) and (6), with wage increases as high as 4.1-4.2 percent. With high school graduates constituting 70 percent of the Venezuelan population in Roraima’s formal labor market, the event studies reveal that these effects emerge post-immigration. This weaker impact for high schoolers might be attributable to the Venezuelan presence in the formal market generating a substitution effect, hampering any external wage growth factors. Yet, the presence of significant positive outcomes suggests that any substitution effects are still likely counterbalanced by other factors. This finding is consistent with \cite{card_impact_1990} and \cite{clemens_labor_2019}, observing no substitution effects on the labor market when analyzing wage level changes due to immigrant influx in lower education individuals. Furthermore, the overall wage increase in the aggregate market implies that immigrants in general primarily serve as a complementary effect more than a substitution effect.

\begin{table}[htb!]
\vspace{0.3in}
\centering
\caption{Heterogeneous Treatment Effects by Education Cohorts}
\label{tab:education}
\begin{threeparttable}

\begin{tabular}[t]{lcccccc}
\toprule
\multicolumn{1}{c}{ } & \multicolumn{2}{c}{College Education} & \multicolumn{2}{c}{High School Education} & \multicolumn{2}{c}{Low Education} \\
  & (1) & (2) & (3) & (4) & (5) & (6)\\
\midrule
Treat: Binary & \num{-0.037}*** &  & \num{0.015}*** &  & \num{0.041}*** & \\
 & (\num{0.006}) &  & (\num{0.004}) &  & (\num{0.012}) & \\
Treat: VZ Ratio x 100 &  & \num{-0.013}*** &  & \num{0.017}*** &  & \num{0.042}***\\
 &  & (\num{0.003}) &  & (\num{0.003}) &  & (\num{0.007})\\
\midrule
Individual FE & X & X & X & X & X & X\\
Year FE & X & X & X & X & X & X\\
\midrule
R2 Adj. & \num{0.886} & \num{0.886} & \num{0.845} & \num{0.845} & \num{0.857} & \num{0.857}\\
RMSE & \num{0.272} & \num{0.272} & \num{0.236} & \num{0.236} & \num{0.181} & \num{0.181}\\
N Clusters & \num{53.000} & \num{53.000} & \num{53.000} & \num{53.000} & \num{53.000} & \num{53.000}\\
N & \num{60575} & \num{60575} & \num{335147} & \num{335147} & \num{176071} & \num{176071}\\
\bottomrule
\end{tabular}

\begin{tablenotes}
\footnotesize 
\item [1] Standard-errors are clustered by municipality. 
\item [2] Propensity score explanatory variables are gender and race indicators, age, age-squared, tenure, and tenure-squared. 
\item [3] * $p < 0.1$, ** $p < 0.05$, *** $p < 0.01$
\end{tablenotes}
\end{threeparttable}

\end{table}

\subsubsection{Heterogeneity by Industry and Occupation}

To investigate further whether immigrants in the formal labor market affected wages, we now focus on the channels of occupation and economic sectors. We create a variable to measure the Venezuelan-to-Brazilian ratio in Roraima across various top 2-digit economic activities in our data. We then divided native workers into two groups: those in sectors with no Venezuelan presence (zero ratio) and those in sectors with Venezuelan presence (positive ratio).

Our preliminary observations indicated that immigrants in RAIS were primarily employed in sectors requiring manual labor, such as retail and wholesale, restaurants, construction, and gardening. Conversely, sectors like finance, insurance, telecommunications, research, pharmaceutics, and entertainment saw little immigrant participation. We applied a similar methodology for occupation analysis, using more detailed 3-digit codes from our dataset. The results of these analyses are presented in Table \ref{tab:act_occ_results}.

Columns (1) and (2) of the table display the outcomes for economic activities and occupations with positive immigrant presence, analyzed using binary and continuous treatment variables. In both cases, we observe an average of around 2 percent wage increase in Roraima post-treatment when a Venezuelan was working in similar occupations or industries. Columns (3) and (4) show results for industries and occupations without any Venezuelan workers recorded in our data. Here, the wage effects were slightly higher for economic activities using the binary treatment version and lower (though less precise) in the continuous treatment version. For occupations, the results showed almost a 3 percent average wage increase.

These findings suggest that the positive effects observed in the aggregate market are mirrored in our economic sector analysis. Specifically, sectors and occupations with a high concentration of immigrants saw a marginally lower average wage increase compared to those without immigrants.

In summary, the entry of immigrants into Roraima appears to have generally driven up wages, likely due to increased labor demand and increased presence of Venezuelan workers in informality. However, those immigrants who integrated into the formal labor market exerted a slight downward pressure on native wages. This effect was not solely based on education levels but also the types of occupations and economic sectors, as evidenced by the immigrants' sorting patterns in the labor market.

\begin{table}[htb!]
\centering
\caption{Heterogeneous Treatment Effects by Activity and Occupation}
\label{tab:act_occ_results}
\begin{threeparttable}

\begin{tabular}[t]{lcccc}
\toprule
 & \multicolumn{2}{c}{Any Immigrants} & \multicolumn{2}{c}{No Immigrants} \\
\cmidrule(l{3pt}r{3pt}){2-3} \cmidrule(l{3pt}r{3pt}){4-5}
  & (1) & (2) & (3) & (4)\\
  \multicolumn{5}{l}{Economic Activity} \\
\midrule
Treat: Binary & \num{0.019}*** &  & \num{0.023}*** & \\
 & (\num{0.005}) &  & (\num{0.007}) & \\
Treat: VZ Ratio x 100 &  & \num{0.019}*** &  & \num{0.014}**\\
 &  & (\num{0.002}) &  & (\num{0.005})\\
 \addlinespace
  \multicolumn{5}{l}{Occupation} \\
\midrule
Treat: Binary & \num{0.022}*** &  & \num{0.027}*** & \\
 & (\num{0.005}) &  & (\num{0.009}) & \\
Treat: VZ Ratio x 100 &  & \num{0.021}*** &  & \num{0.030}***\\
 &  & (\num{0.003}) &  & (\num{0.005})\\
\midrule
Individual FE & X & X & X & X\\
Year FE & X & X & X & X\\
\midrule
R2 Adj. & \num{0.855} & \num{0.855} & \num{0.931} & \num{0.931}\\
RMSE & \num{0.241} & \num{0.241} & \num{0.200} & \num{0.200}\\
N Clusters & \num{53} & \num{53} & \num{53} & \num{53}\\
N & \num{534212} & \num{534212} & \num{34093} & \num{34093}\\
\bottomrule
\end{tabular}

\begin{tablenotes}
\footnotesize 
\item [1] Standard-errors are clustered by municipality. 
\item [2] Propensity score explanatory variables are gender and race indicators, age, age-squared, tenure, tenure-squared and education level. 
\item [3] * $p < 0.1$, ** $p < 0.05$, *** $p < 0.01$
\end{tablenotes}
\end{threeparttable}

\end{table}

\subsubsection{Substitution through Occupation Changes of Brazilian Workers}

To establish more robustness to our assumption of small but significant substitution effects happening in the formal sector, we can examine the job changes of native workers in response to immigrant sorting. This is similar to the approach taken in \cite{foged_immigrants_2016}, which assumes that refugees entering the labor market took on manual-intensive jobs, potentially allowing native workers to shift into other occupations.

To analyze the effects of immigration on native workers' job choices, we create a binary variable to indicate whether an individual in occupations where immigrants are observed ever changed their occupation to a position where we do not observe Venezuelans. Similar to previous frameworks, we compare these occupational changes across the control and treatment groups. 

We use this variable as the dependent variable in Equation (\ref{eq:wagesdid_bin}) and (\ref{eq:wagesdid_cont}), presenting the regression results in Table \ref{tab:mover}. They correspond to a linear probability measurement of an individual being a ``mover'' from immigrant occupations to non-immigrant ones.

\begin{table}[htb!]
\centering
\caption{Movement from Occupations with Any Immigrants to Occupations with No Immigrants}
\label{tab:mover}
\begin{threeparttable}

\begin{tabular}[t]{lcc}
\toprule
\multicolumn{1}{c}{ } & Movement & Movement \\
  & (1) & (2)\\
\midrule
Treat: Binary & \num{0.002}*** & \\
 & (\num{0.000}) & \\
Treat: VZ Ratio x 100 &  & \num{0.001}***\\
 &  & (\num{0.000})\\
\midrule
Individual FE & X & X\\
Year FE & X & X\\
\midrule
R2 Adj. & \num{0.008} & \num{0.008}\\
RMSE & \num{0.072} & \num{0.072}\\
N Clusters & \num{53.000} & \num{53.000}\\
N & \num{576800} & \num{576800}\\
\bottomrule
\end{tabular}

\begin{tablenotes}
\footnotesize 
\item [1] Standard-errors are clustered by municipality. 
\item [2] Propensity score explanatory variables are gender and race indicators, age, age-squared, tenure, tenure-squared and education level. 
\item [3] * $p < 0.1$, ** $p < 0.05$, *** $p < 0.01$
\end{tablenotes}
\end{threeparttable}

\end{table}

The results indicate that, overall, there was a small but significant change of occupation for Brazilians conditional on the presence of Venezuelans. This implies that even though the substitution effect is measurable, it should be considered too small to provide any significant overall effect in the market.

\subsection{Relationship between Formal Sector Effects and Informal Sector Effects}

So far, our analysis has demonstrated the positive labor market effects of the Venezuelan immigration crisis. We also showed that individuals not involved in occupations or economic sectors and occupations with a sizeable Venezuelan presence experienced more prominent wage increases compared to their counterparts. Moreover, we did not observe any negative wage effects for Brazilians in manual labor occupations where immigrants were mostly concentrated.

One way to interpret these results is to consider the role of Venezuelan immigrants who are not part of the formal market. We showed that many Venezuelan immigrants in the region had sought refugee status, but only a small percentage entered the formal market. This potentially implies that many of them worked informally or were in seek of employment. Informal workers tend to have lower levels of education and they specialize in manual tasks, while those in the formal labor market often hold cognitive or technical jobs, with some overlap.

In Roraima, about 45 percent of the workforce is involved in the informal labor market. To analyze whether there are negative impacts on wages in the informal labor market, we use the PNAD-C dataset from 2012-19. PNAD-C is a representative household survey conducted by the Brazilian Institute of Geography and Statistics (IBGE) every quarter that includes a variety of socioeconomic information, such as employment status, income, race, gender, and education level among others. However, a major limitation of the dataset is that, unlike RAIS, social identification of firms and persons, and nationality, are not observable. Therefore, we cannot directly analyze the impacts on Brazilian citizens or use individual fixed effects to control for time-invariant individual characteristics. Moreover, unlike RAIS, we cannot observe the municipality of individuals. As a result, estimates from PNAD-C are potentially downward biased. The adapted model is specified as follows:
\begin{equation} \label{eq:pnad_did}
    y_{ist} = \beta D_{st} + f(X_{it}) + \theta_s + \alpha_t + \varepsilon_{ist}
\end{equation}

\noindent
where $y_{ist}$ represents the logarithmic average wage that individual $i$ earned in quarter $t$ in state $s$. $D_{ist}$ is the indicator function that becomes one if individual $i$ is from the state of Roraima after 2013. $f(X_{it})$ is the covariate matrix linear function, including education, race, gender, age, and age-squared. State fixed effects and year fixed effects are represented by $\theta_s$ and $\alpha_t$, respectively. The term $\beta$ estimates the effects of the Venezuelan refugee crisis in Roraima on logged wages. The error term $\varepsilon_{ist}$ is clustered at the state level, given we cannot observe municipality in the PNAD-C data.

\begin{table}[htb!]
\centering
\caption{Log Wage Effects in the Informal Sector}
\label{tab:pnadc_reg}
\begin{threeparttable}

\begin{tabular}[t]{lccc}
\toprule
 & \multicolumn{3}{c}{Informal Log Wages} \\
  \cmidrule(l{3pt}r{3pt}){2-4} \\
  & (1) & (2) & (3) \\
\midrule
Treat  & \num{-0.018} & \num{-0.022} & \num{0.012}\\
 & (\num{0.029}) & (\num{0.027}) & (\num{0.029})\\
 \addlinespace
Immigrant Occupations & &  \num{0.077}*** & \\
 & & (\num{0.016}) &  \\
Treat × Immigrant Occupations & & \num{0.011} &  \\
 & & (\num{0.019}) & \\
 \addlinespace
Immigrant Activities &  & & \num{0.140}*** \\
 &  &  & (\num{0.012})\\
Treat × Immigrant Activities &  & & \num{-0.085}*** \\
 & &  & (\num{0.017}) \\
\midrule
N & \num{67894} & \num{67894} & \num{67894}\\
Year FE & X & X &X \\
State FE &X &X &X\\
\bottomrule
\end{tabular}
\begin{tablenotes}
\footnotesize 
\item [1] Standard-errors are clustered by state. 
\item [2] Propensity score explanatory variables and covariates in the main spacification are gender and race indicators, age, age-squared, tenure, tenure-squared and education level. 
\item [3] * $p < 0.1$, ** $p < 0.05$, *** $p < 0.01$
\end{tablenotes}
\end{threeparttable}

\end{table}

Table \ref{tab:pnadc_reg} presents the outcomes of our analysis using equation (\ref{eq:pnad_did}). Column (1) shows the aggregate treatment effect result, at around -1.8 percent, which is not statistically significant. Despite the inherent coarseness of the PNAD-C survey data, it still provides limited evidence of a substitution effect in the informal market due to the increased labor supply in Roraima.

Moreover, PNAD-C includes data on economic activities and worker occupations, albeit at a lower level of detail. To assess the impact of immigrant exposure on specific industries or job occupations more accurately, we link sectors with a higher immigrant presence in both variables in RAIS to corresponding categories in PNAD-C. We assume that Venezuelan immigrants have similar preferences regarding economic activities and occupations in both informal and formal sectors. The categories used for the indicator function are general service, crafting, and construction helpers for job occupations, and construction, restaurants, and commerce for economic activities.

The model, similar to Equation (\ref{eq:pnad_did}), differs by allowing the heterogeneity dummy to interact with the treatment indicator. Columns (2) and (3) in Table \ref{tab:pnadc_reg} show results for the informal sector; specifically, the interactions with occupations and economic activities, respectively.

In the informal sector, the data indicate that both occupations and firm activities were associated with higher wages compared to other categories. However, with the treatment interaction, the results become statistically insignificant for occupations and show a significant 8.5 percent decrease for firm activities. This suggests that immigrants may have adversely affected the earnings of individuals in these categories in Roraima post-crisis.

The results indicate a significant impact of immigration on the informal sector, highlighting a substitution effect that adversely influences wages. Conversely, in the formal sector, the analysis suggests a potential benefit for workers, attributable to the complementarity effect that the informal sector provides to formal employment and a possible increase in efficiency within the formal market.

\section{Conclusion} \label{sec:conclusion}

In this paper, we conducted a comprehensive analysis of the labor market impacts of the Venezuelan crisis in Brazil, focusing on the state of Roraima, where the crisis had a direct impact. The state's geographical isolation helps us use the immigration shock as a natural experiment, allowing us to measure its effects on the local labor market. Using a difference-in-differences model, we explored the potential differences in effects based on market diversity in terms of economic sector and worker occupation.

Our findings reveal that the average monthly wages in Roraima increased by approximately 2 percent in the early stages of the crisis. By analyzing the presence of Venezuelan formal workers by firm's economic sectors and occupation type, we found that Brazilian workers involved in economic sectors and occupations that had a presence of Venezuelan immigrants experienced a slightly lower wage increase than those involved in sectors and occupations without any Venezuelan presence. We also observed no significant changes in formal employment displacement of Brazilians but did find evidence of them moving from high-immigrant occupations to low-immigrant occupations in the post-treatment years.

Furthermore, using nationally representative survey data, our analysis of the informal labor market revealed that while Brazilian informal workers did not experience a significant drop in wages on aggregate, those involved in sectors with the presence of Venezuelan immigrants experienced a significant wage drop. We can conclude that immigrants in the informal labor market acted as complements to the formal labor market, allowing the overall formal wage to increase and offsetting any substitution effect of foreign workers within formality.

In summary, our study emphasized the need to consider the various factors that can influence the impacts of refugees on the labor market. While our research suggested that refugees can bring benefits, it also highlighted the potential drawbacks of large-scale immigration in regions with a significant informal economy. Moving forward, policymakers should prioritize policies that improve the welfare of refugees and promote their active participation in the economy while also being mindful of the potential negative effects on those working informally. Future research should further explore the impacts of the population boom in Roraima, including improving the understanding of the effects on consumption, health, and public safety outcomes.

\clearpage
\setlength\bibsep{0pt}
\bibliographystyle{econ}
\small \bibliography{references}

\normalsize

\clearpage


\appendix
\section*{Appendices}
\counterwithin{figure}{section}
\counterwithin{table}{section}







\section{Foreign Presence Outside RAIS} \label{sec:vz_permanence_in_rr}
Even though RAIS provides a good picture of Venezuelans entering Brazil and exclusively staying in Roraima, it only counts Venezuelans in the formal labor market. Venezuelans may be crossing the border and going through, staying in other states outside the formal labor market, in refugee camps, or working informally. The Federal Police data shows that around 41 thousand individuals crossed Roraima and did not return to Venezuela. However, we do not observe either in RAIS or the Federal Police data whether they stayed in Roraima or moved around.

It would jeopardize our identification strategy if the control states hosted many Venezuelan refugees outside the formal labor market. To show that it is not the case, we rely on the refugee application data from the Brazilian National Committee for Refugees (CONARE).

Foreigners in Brazil can be registered as refugees to get benefits such as obtaining the individual taxpayer registration number (CPF), accessing health and education services, and opening a bank account, among others. A potential refugee must submit its recognition to, and then analyzed by, CONARE. The committee then decides whether they are recognized as a refugee. If rejected, they can appeal. Since refugee applications only exist conditional on the presence of a forcibly displaced population, we believe the data is an adequate proxy for Venezuelans not observed in the formal labor market. 

Variables included in CONARE are the nationality of the applicants, the reason for leaving their country, the date when the application was submitted, the municipality and the state where the application was submitted, and the date when CONARE made the decision.

\begin{figure}[htb!]
    \centering
    \includegraphics[width =0.8\textwidth]{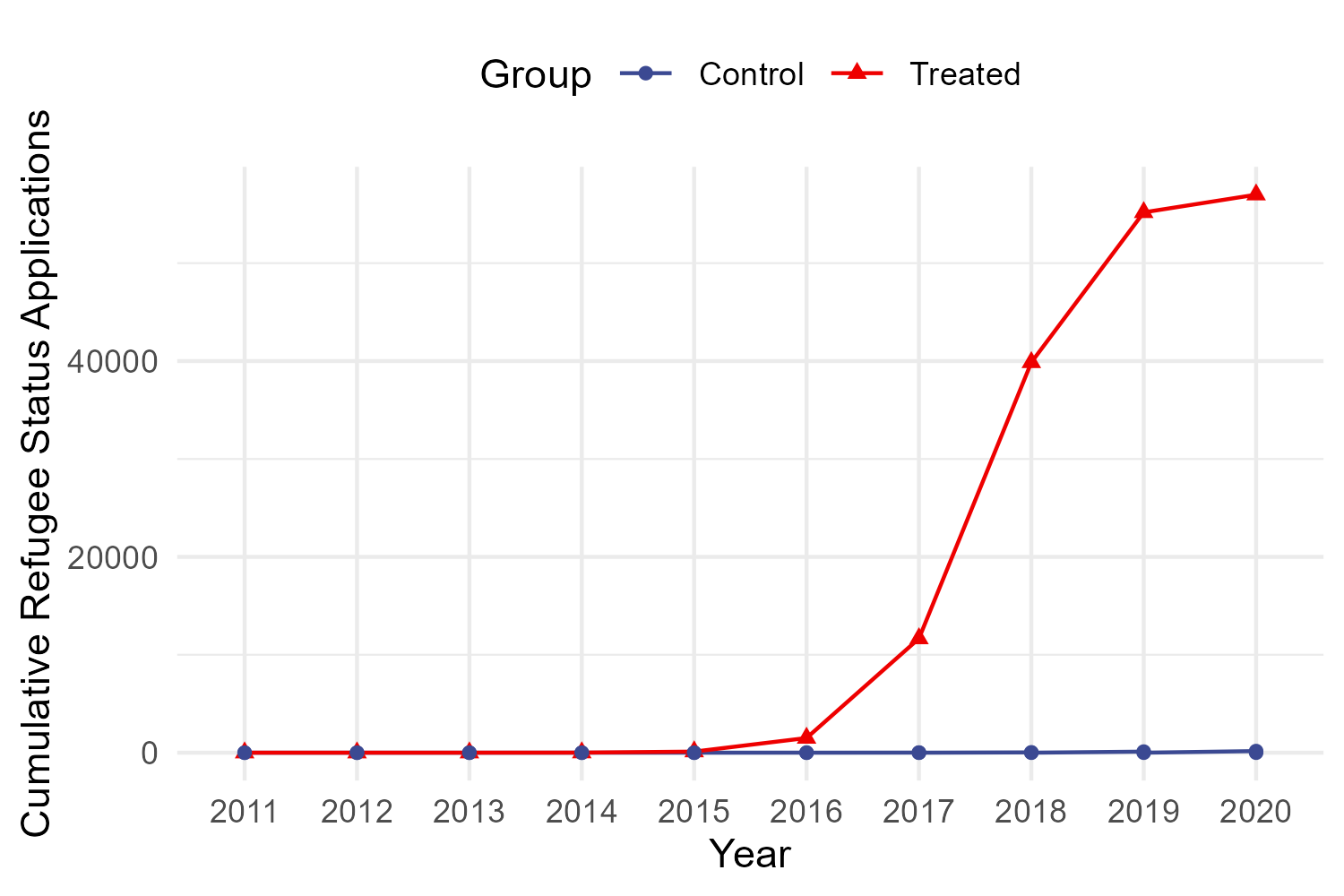}
    \caption{Cumulative refugee requests by year and treatment status}
    \label{fig:conare_trend}
\end{figure}

Table \ref{fig:conare_trend} shows the cumulative number of refugee applications by treatment status and year between 2011 and 2020. There were 56,984 refugee requests in Roraima in 10 years, with the first application submitted in 2015. If we consider our period of interest, 2014-2017, more than 10 thousand individuals requested refugee status, with virtually zero applications found in control states. If immigrants are moving across treated and control states, the likelihood of a Venezuelan applying for refugee status in another location rather than Roraima would significantly increase. Accordingly, we do not see this behavior in the data.

Another consideration is that only a quarter of those who entered Roraima and stayed in 2017 applied for refugee status. If we combine RAIS, around ten percent of these applicants found a job in the formal labor market. Suppose we assume an individual seeks refugee status to apply for employment and proper residence. In that case, it is a safe inference that a significant fraction of Venezuelans were actively in seek of employment, not necessarily in the formal labor market, with some of these accepting offers in the informal labor market.






\clearpage
\section{Synthetic Control Methods Results} \label{sec:scm_appendix}

Concerns may arise in our estimation strategy of using the doubly robust difference-in-differences method given the nature of our natural experiment with only one treatment state (Roraima) if the control group is chosen incorrectly. In the main analysis, we compare the treatment state with carefully chosen control states in Brazil with states with socioeconomic and sociodemographic characteristics comparable to those of the treatment state. We also showed pre-treatment parallel trends in our outcome variable. However, it may lead to endogeneity bias if the control groups are characteristically different from the treatment state.

A popular way of estimating the causal effects of the treatment in the case of a singular or a few treatment states is by using the Synthetic Control Method (SCM). Using SCM with the northern Brazilian states as our donor pool for the synthetic Roraima, we reconfirm the validity of our choice of control states in the main analysis.

Consider the case with $j+1$ states where $j=0,...,J$ where $j=0$ is the only treatment unit, in our case, the state of Roraima. $W$ is the $j \times 1$ vector of non-negative weights for the donor states that comprise synthetic Roraima such that $\sum^{1}_{J} w_{j} = 1$ and $0 \leq w_{j} \leq 1$. $X_1$ is a $K \times 1$ vector of matching variables used to construct our synthetic control group. In our case, our main primary outcome variables, the wages, act as the only matching variable. $X_0$ is a $K \times J$ matrix for all control states J with the same predictors as Roraima's pretreatment wage trends. The SCM selects the vector of weights $W*$ that minimizes the difference between $X_1$ and $X_{0}W$ such that:
\begin{align}
W^{*} &= \argmin_{w} (X_{1} - X_{0}W)'(X_{1}-X_{0}W) \\ 
& \text{s.t.} \sum^{j-1}_{J}w_{j}=1, 0 \leq w_{j} \leq 1
\end{align}

The optimal $W$ is chosen by an optimization process that minimizes the mean-squared prediction error (MSPE) in the pre-treatment period. Following the construction of weights, we estimate the effects of the wage effects of the Venezuelan refugee crisis in Roraima, Brazil using the formula: $(Y^{RR}_{Post} - Y^{RR}_{Pre}) - (Y^{SRR}_{Post} - Y^{SRR}_{Pre})$, where $Y^{RR}_{Pre}$ and $Y^{RR}_{Post}$ are the average logged wage in Roraima in pre- and post-treatment periods, and $Y^{SRR}_{Pre}$ and $Y^{SRR}_{Post}$ are the average logged wage in synthetic Roraima in pre- and post-treatment periods respectively.

\begin{figure}[htb!]
    \centering
\includegraphics[width=0.6\textwidth]{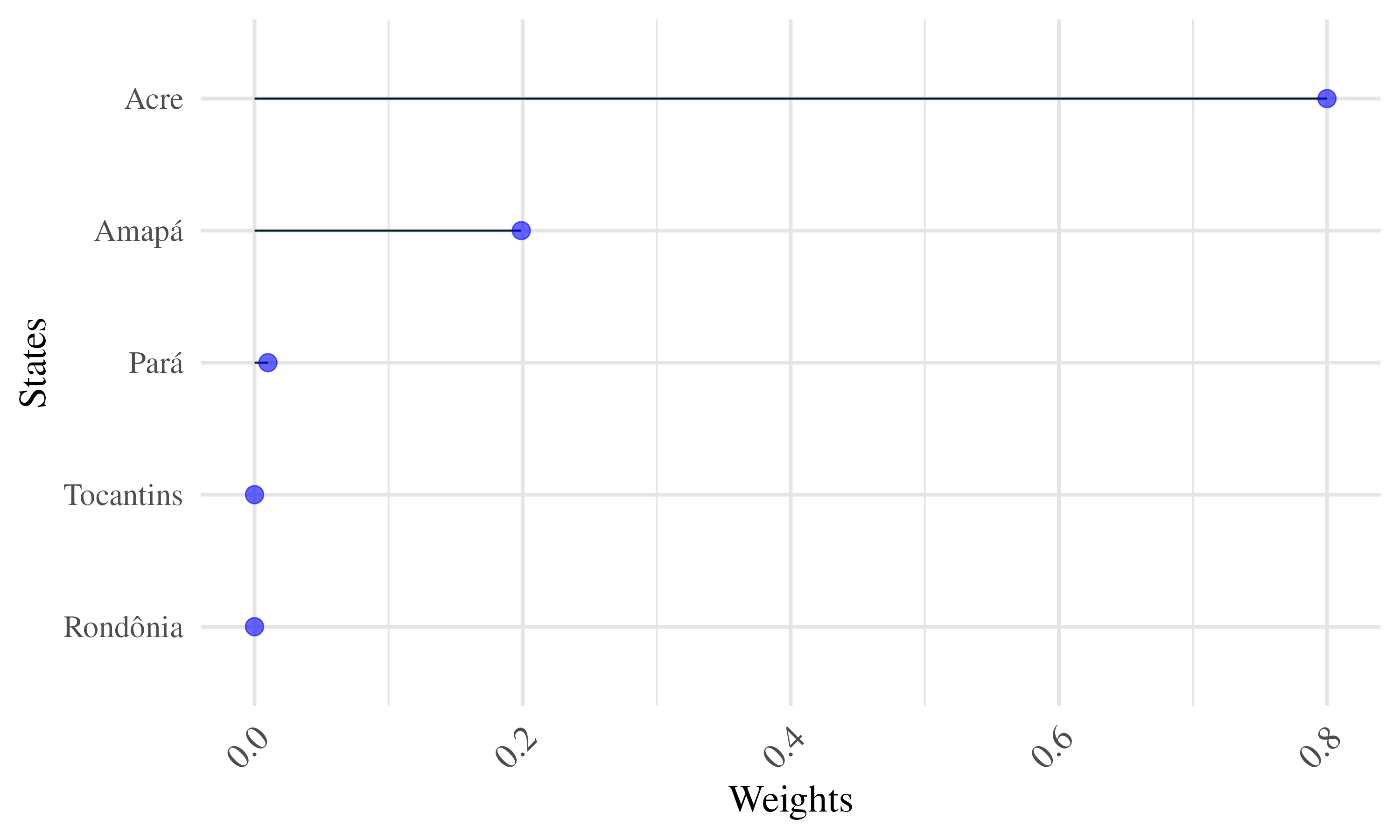}  
        \caption{Weights for Donor States in the Northern Brazilian Region in the Synthetic Control Method}
\label{fig:fig_scm_weights}
\end{figure}

Due to the geographical proximity and sociodemographic similarities with Roraima, our donor pool consists of northern Brazilian states. Figure \ref{fig:fig_scm_weights} illustrates the weights we received from the optimization procedure. The entire weight is shared between Acre (0.80) and Amapa (0.20), which confirms our initial decision to use these states as the control states in our main analysis.

\begin{figure}[htb!]
    \centering
        \includegraphics[width=0.8\textwidth]{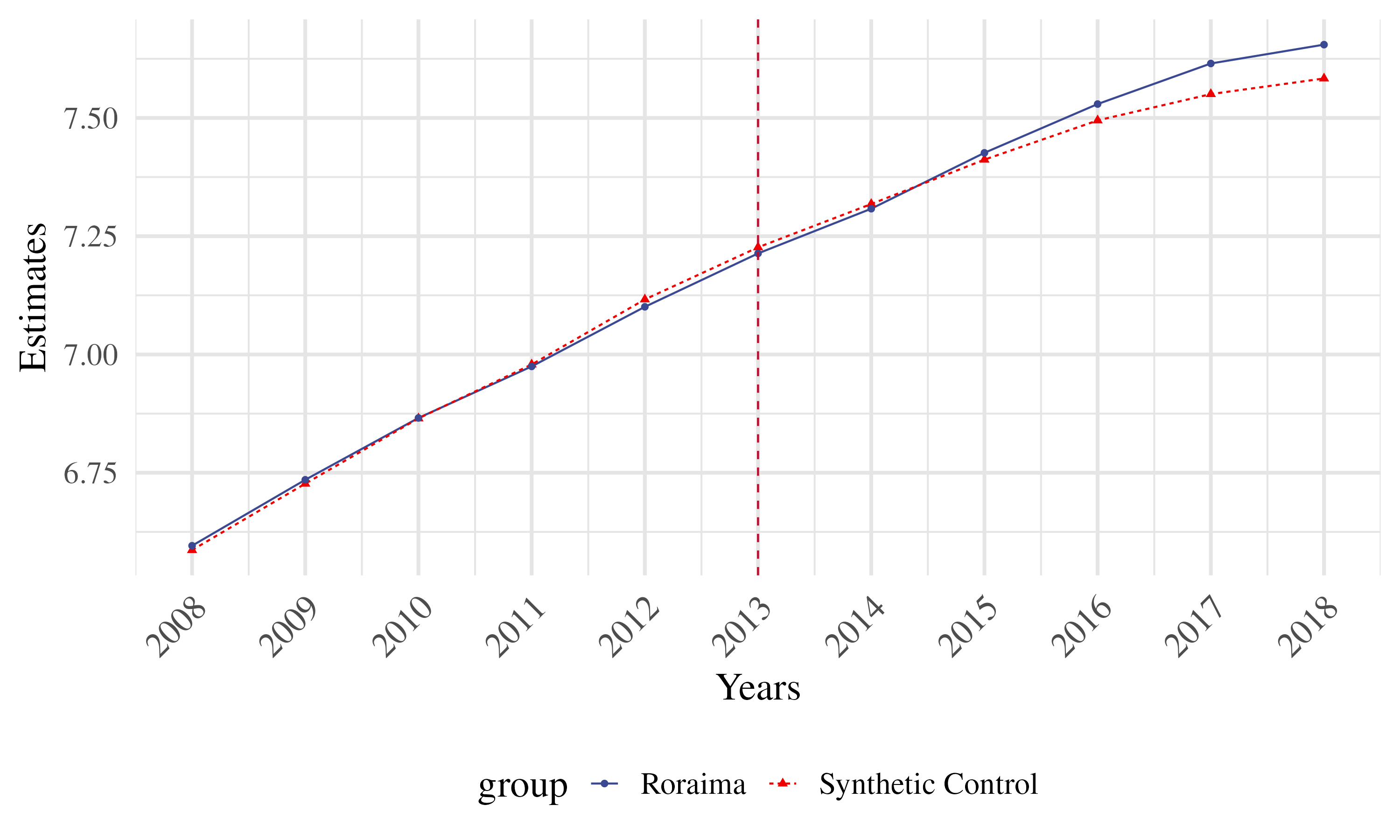}  
        \caption{Effects of the Venezuelan Refugee Crisis in the Brazilian Labor Market using the Synthetic Control Methods}
        \label{fig:fig_scm_lw}
\end{figure}

Figure \ref{fig:fig_scm_lw} shows the wage results from the SCM analysis. The figure shows perfect parallel trends between Roraima and synthetic Roraima until 2013 and a few years following. The trends start to diverge in 2016 where the wages for synthetic Roraima have a reduction in slope but Roraima's slope reduction is less steep. In the year 2018, we see an almost 0.20 percentage point difference between Roraima and synthetic Roraima. This exercise confirms the results from our primary analysis using difference-in-differences.

\clearpage
\section{Synthetic Difference-in-Differences}
\label{sec:sdid}

In this section, we provide results using the Synthetic Difference-in-differences (SDID), approach from \cite{arkhangelsky_synthetic_2021}, a twist from the classical SCM that relaxes its restrictions by imposing only slope-fitting of the synthetic control and the treatment group.

Figure \ref{fig:fig_sdid_weights} shows the optimal donors for the SDID algorithm. It confirms that Amapá and Acre are the most suitable donors for the experiment, with a combined weight value of approximately 65 percent. 

\begin{figure}[htb!]
    \centering
    \vspace{1cm}
\includegraphics[width=0.6\textwidth]{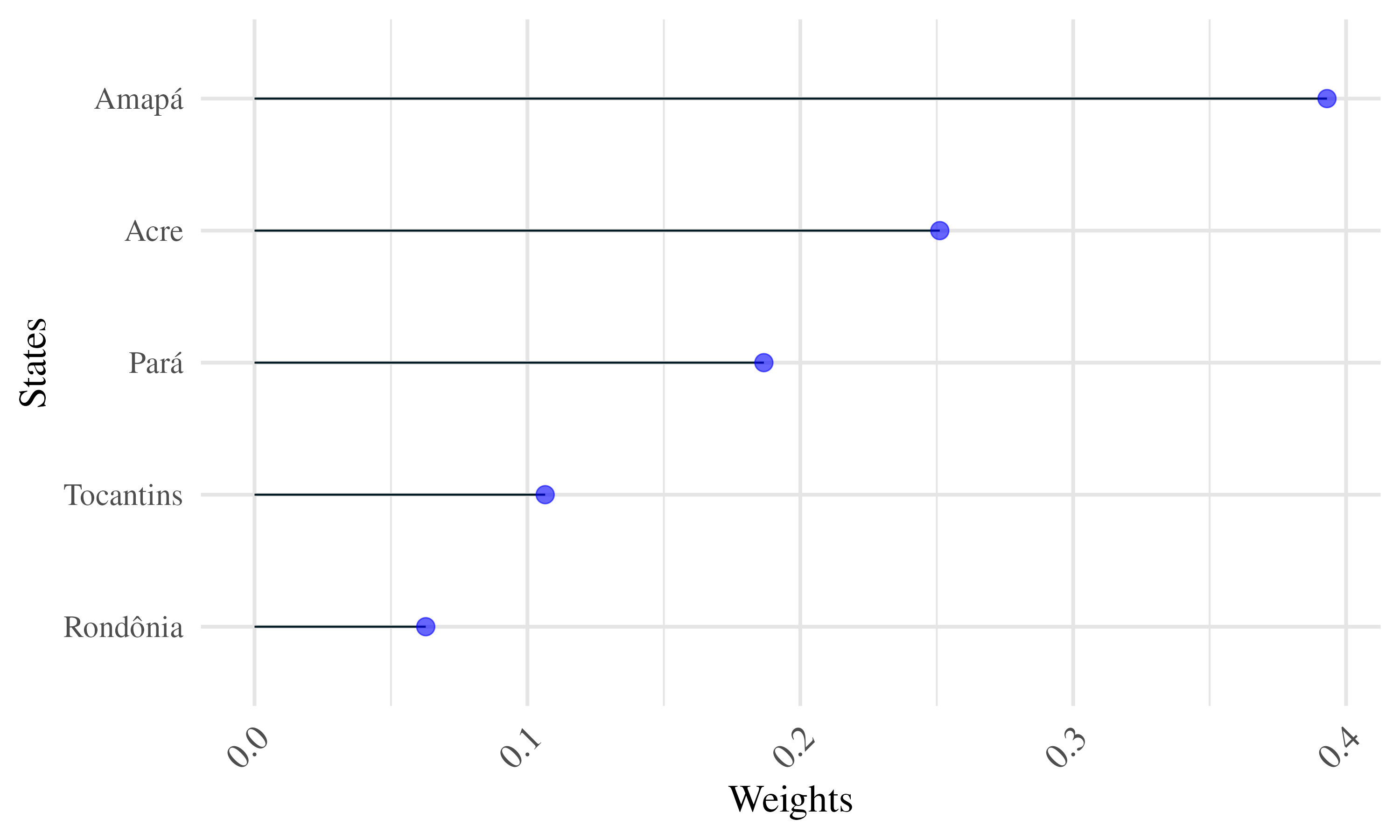}  
        \caption{Weights for Donor States in the Northern Brazilian Region in the Synthetic Difference-in-Differences Method}
\label{fig:fig_sdid_weights}
\end{figure}

Figure \ref{fig:fig_sdid_weights} shows the treatment group and synthetic Roraima trends using the SDID framework. After 2013, Roraima was able to maintain higher wage levels compared to the generated synthetic Roraima, confirming our main results using a straightforward difference-in-differences approach.

\begin{figure}[htb!]
    \centering
        \includegraphics[width=0.8\textwidth]{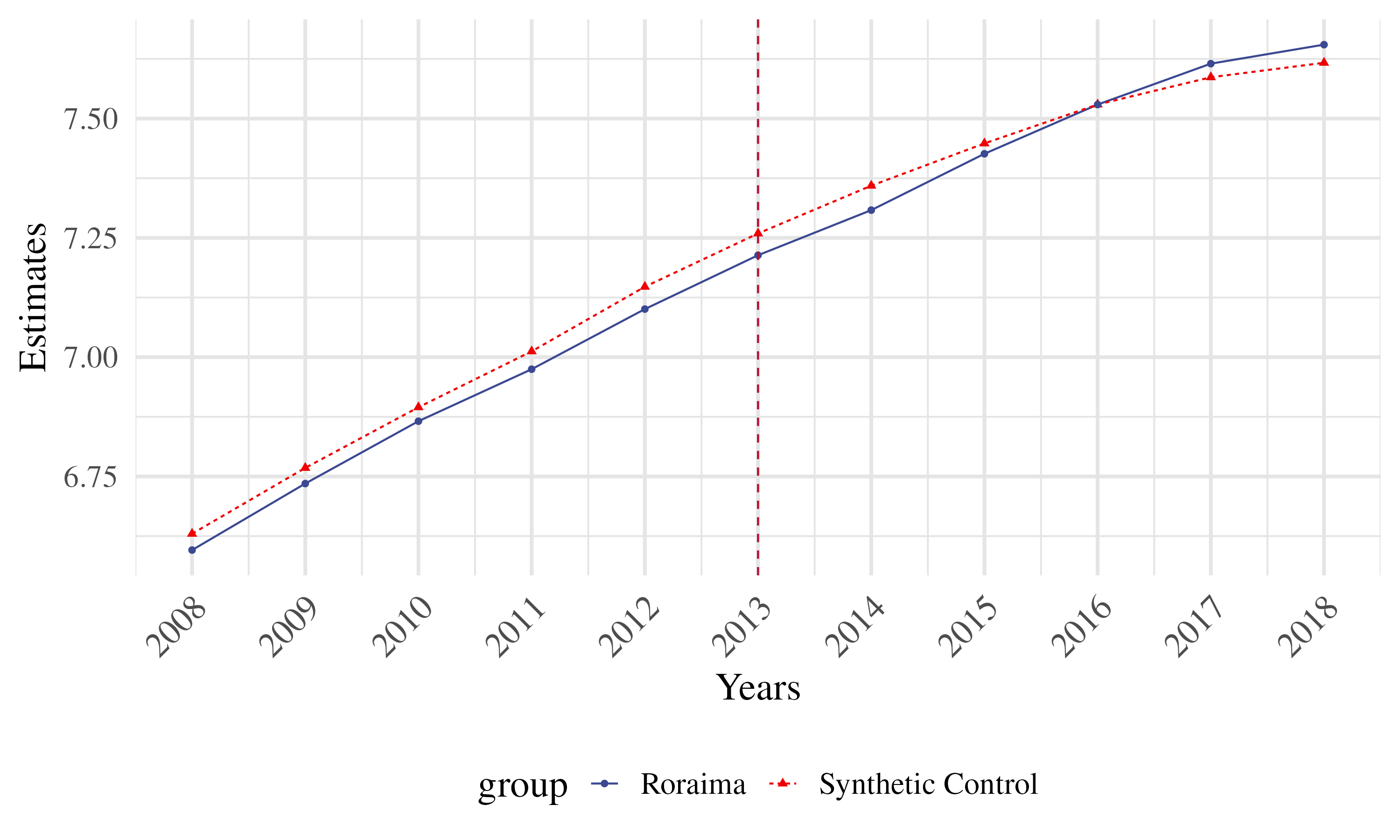}  
        \caption{Effects of the Venezuelan Refugee Crisis in the Brazilian Labor Market using the Synthetic Difference-in-Differences Methods}
        \label{fig:fig_sdid_weights}
\end{figure}

\clearpage 
\section{Other Figures}

\begin{figure}[htb!]%
    \vspace{1cm}
    \centering    
    \caption{Industries where Venezuelan migrants worked in 2018 (as a percentage of total Venezuelans in RAIS)}%
    \includegraphics[width=0.8\textwidth]{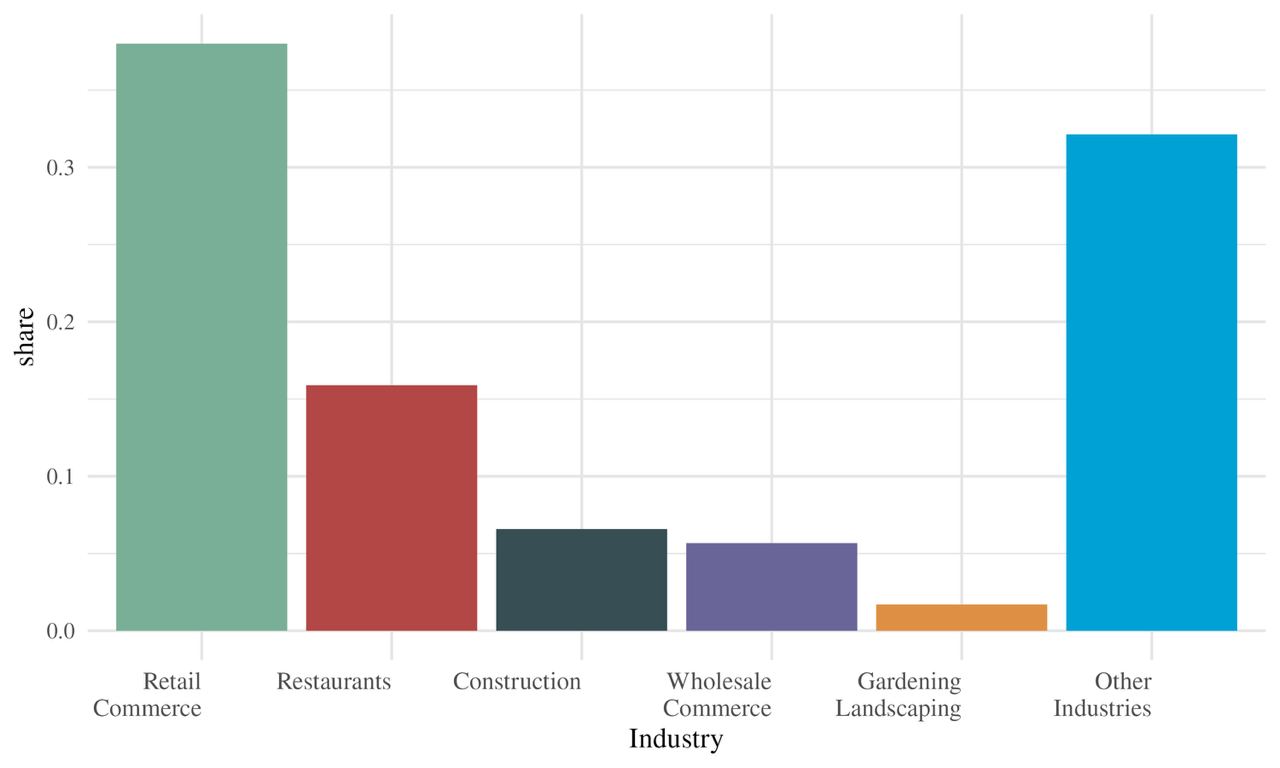}
    \label{fig:industries_vz}%
\end{figure}

\begin{figure}[htb!]%
    \vspace{1cm}
    \centering    
    \caption{Occupations of Venezuelans in 2018 (as a percentage of total Venezuelans in RAIS)}%
    \includegraphics[width=0.8\textwidth]{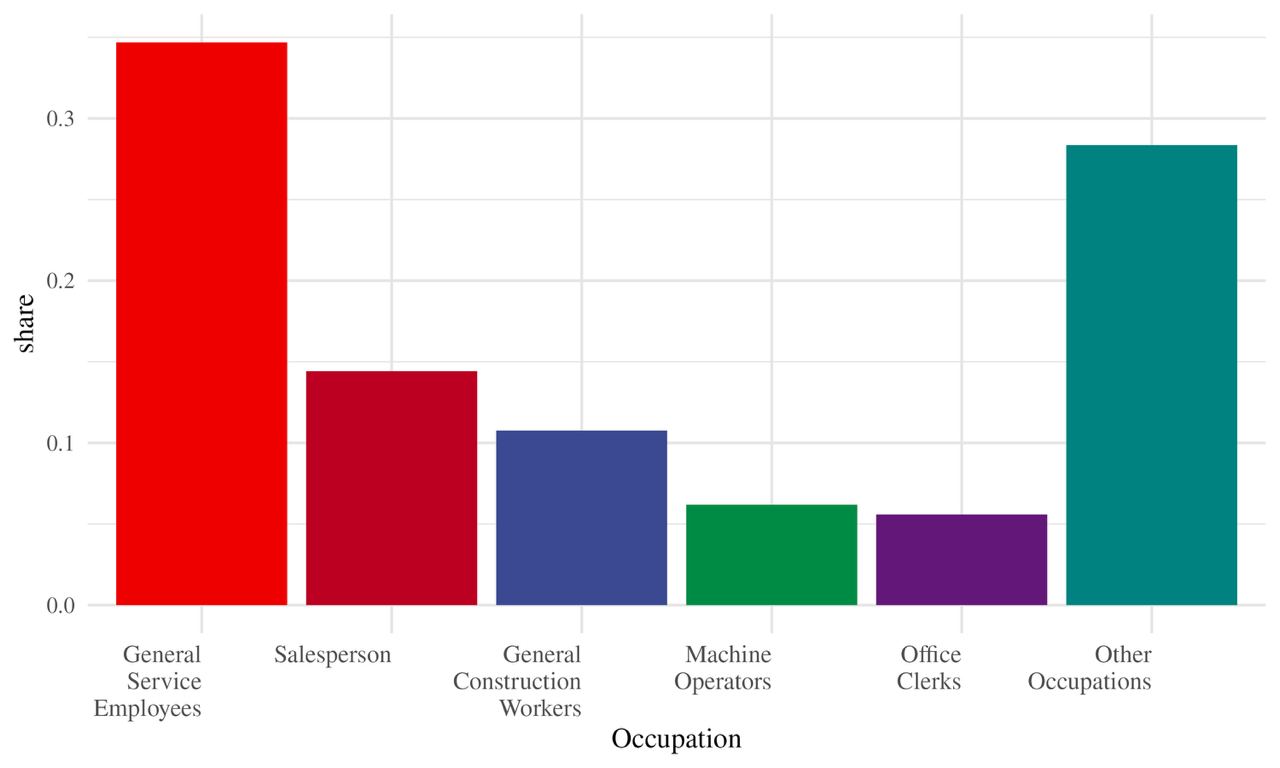}
    \label{fig:occupations_vz}%
\end{figure}

\begin{figure}[htb!]
    \centering
        \includegraphics[width=0.6\textwidth]{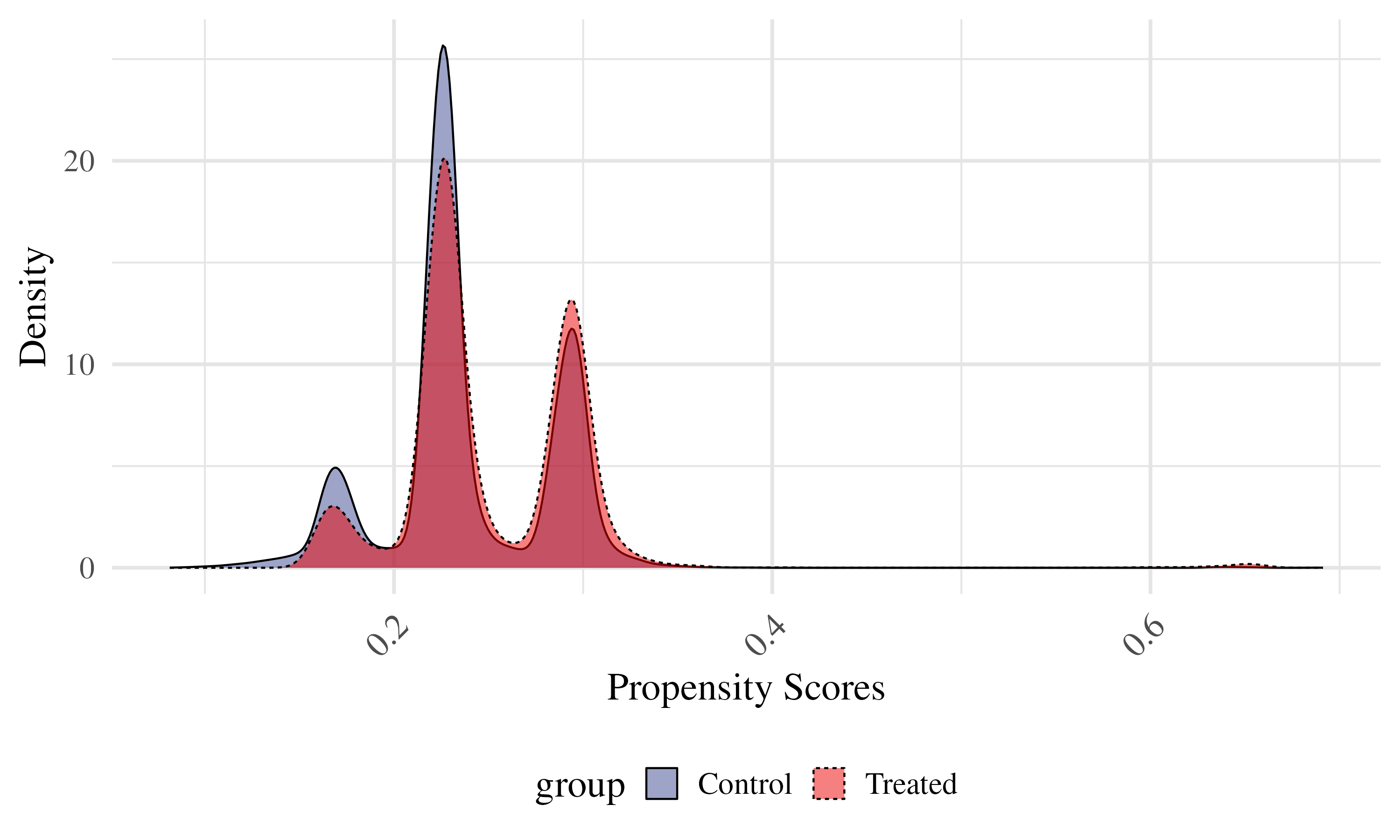}  
        \caption{Covariate Balance for Propensity Score Analysis for College Graduate RAIS Sample}
\label{fig:fig_ps_college_density}
\end{figure}

\begin{figure}[htb!]
    \centering   \includegraphics[width=0.6\textwidth]{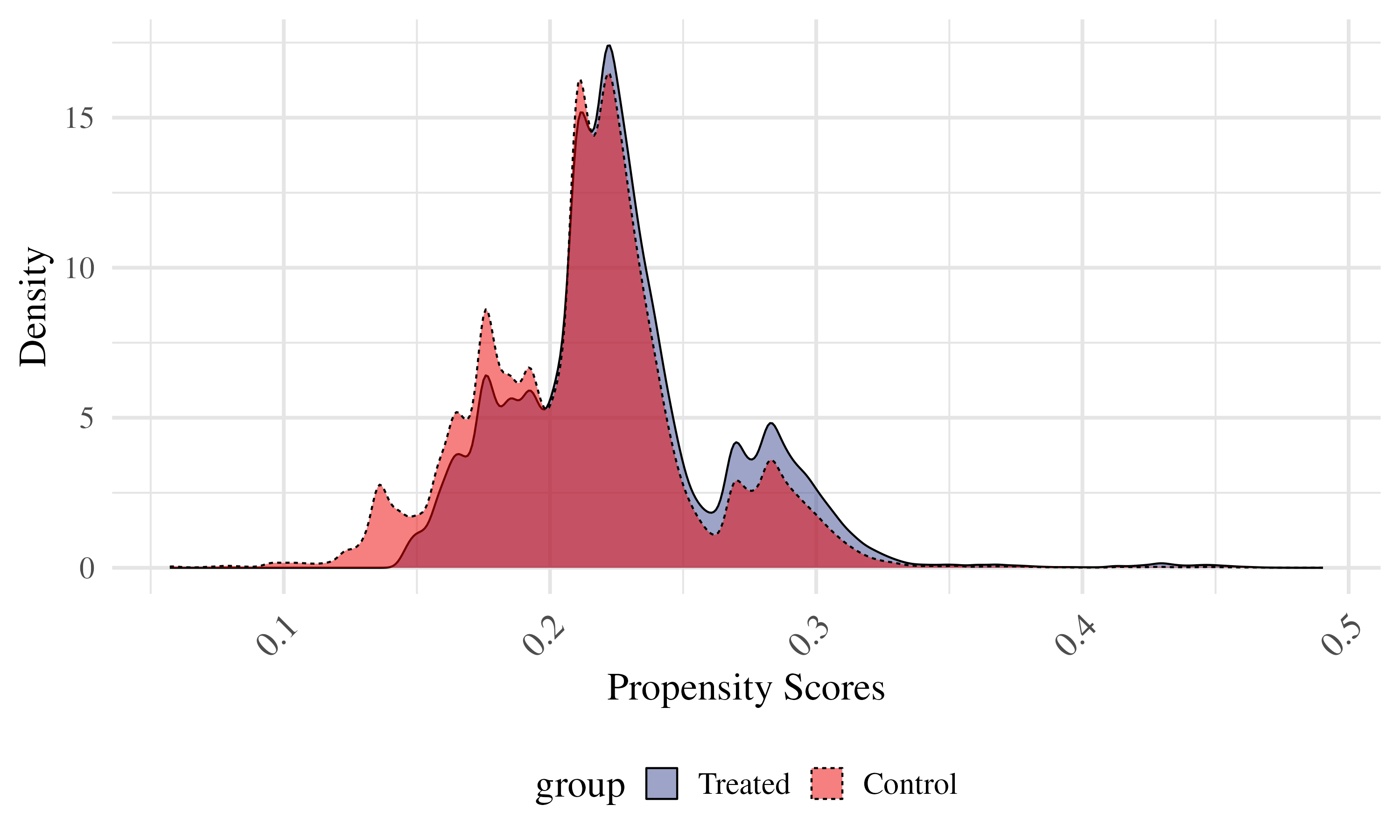}  
        \caption{Covariate Balance for Propensity Score Analysis}
\label{fig:fig_ps_density}
\end{figure}

\begin{figure}[htb!]
    \centering
\includegraphics[width=0.6\textwidth]{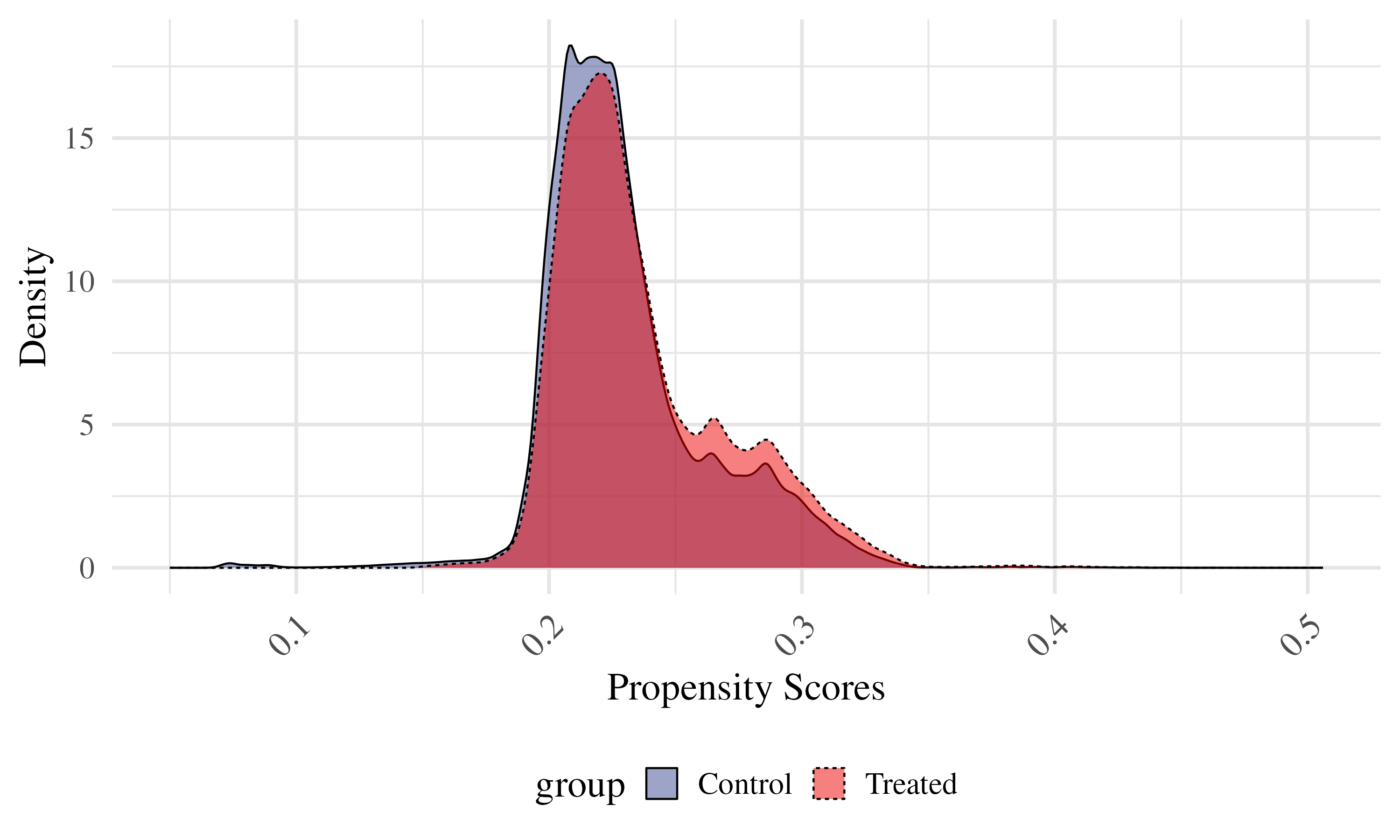}  
        \caption{Covariate Balance for Propensity Score Analysis for High School Graduate RAIS Sample}
        \label{fig:fig_ps_hschool_density}
\end{figure}

\begin{figure}[htb!]
    \centering
        \includegraphics[width=0.6\textwidth]{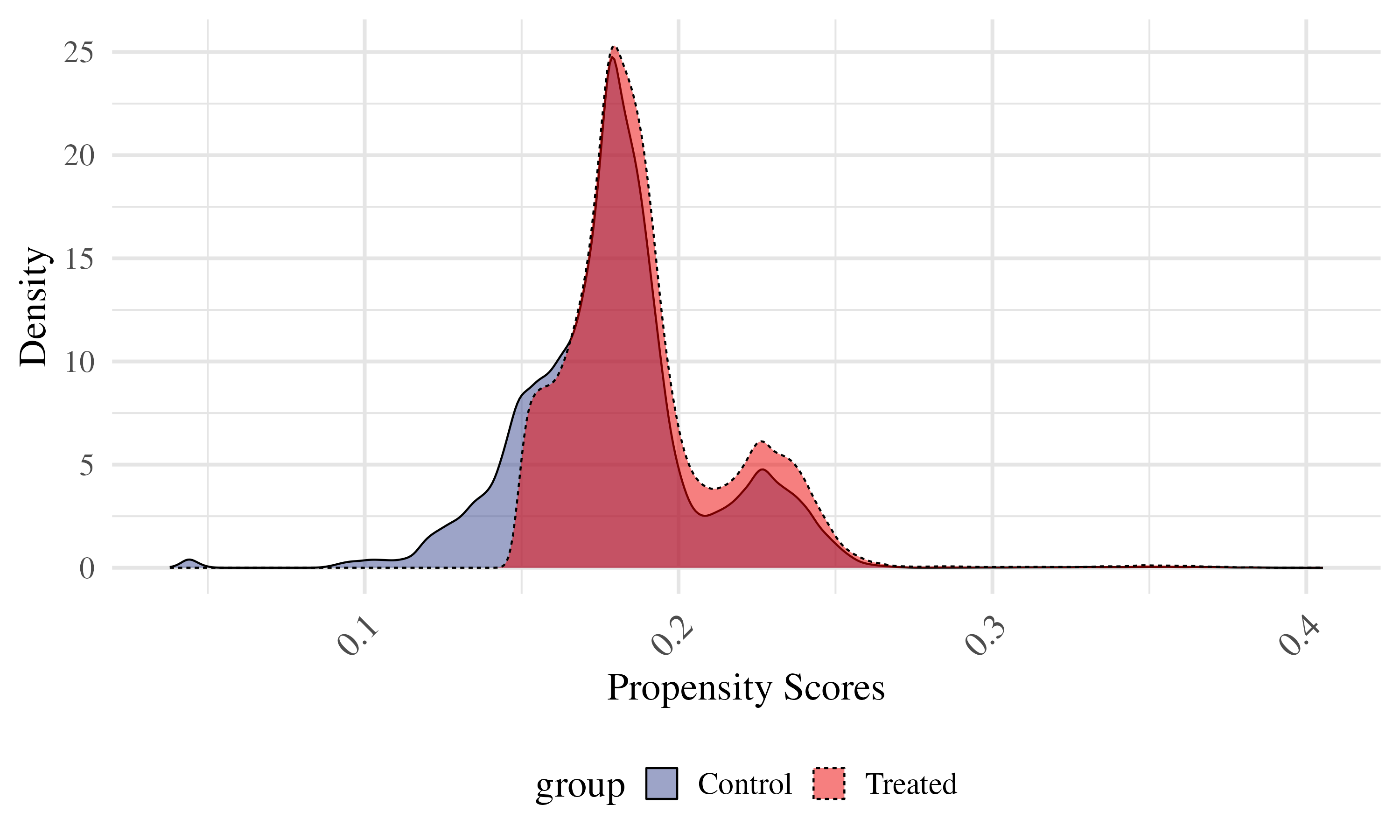}  
        \caption{Covariate Balance for Propensity Score Analysis for RAIS Sample with Less Than High School Education}
\label{fig:fig_ps_low_density}
\end{figure}

\end{document}